\documentclass[reqno,12pt]{article}
\usepackage{ae} 
\usepackage[T1]{fontenc}
\usepackage[ansinew]{inputenc}
\usepackage{amsmath}
\usepackage{amsfonts}
\usepackage{amssymb}
\usepackage{graphicx}
\graphicspath{{Pictures/}}
\usepackage{bm}
\usepackage{float}
\usepackage{color}
\definecolor{darkblue}{cmyk}{0.9,0.9,0,0}
\usepackage[colorlinks=true,linkcolor=darkblue,citecolor=darkblue,urlcolor=darkblue]{hyperref}
\usepackage{epsfig}

\usepackage[dvipsnames]{xcolor}
\usepackage{graphicx}

\usepackage{diagbox}

\usepackage{indentfirst}

\newcommand{\bX}{\bar{X}}
\newcommand{\bY}{\bar{Y}}
\newcommand{\bZ}{\bar{Z}}

\newcommand{\beq}{\begin{equation}}
\newcommand{\eeq}{\end{equation}}
\newcommand\beqa{\begin{eqnarray}}
\newcommand\eeqa{\end{eqnarray}}
\newcommand\bea{\begin{array}}
\newcommand\eea{\end{array}}

\def\XXint#1#2#3{{\setbox0=\hbox{$#1{#2#3}{\int}$}
\vcenter{\hbox{$#2#3$}}\kern-.5\wd0}}

\newcommand{\neqa}{\nonumber\end{eqnarray}}

\newcommand{\p}{\partial}

\def\tr{{\rm tr~}}

\newcommand{\Tr}{{\rm Tr}}

\renewcommand{\d}{\partial}

\newcommand{\<}{{\langle}}
\renewcommand{\>}{{\rangle}}

\newcommand{\re}{\relax{\rm I\kern-.18em R}}

\renewcommand{\sp}{p\hspace{-.40em}/}

\def\su2{{SU(2)}}

\def\[{\left[}
\def\]{\right]}

\def\<{\langle}
\def\>{\rangle}

\def\i2{\frac{i}{2}}

\def\spi{\relax{\rm \pi\kern-0.5em /}}
\def\sA{\relax{\rm A\kern-0.5em /}}
\def\sp{\relax{\rm p\kern-0.5em /}}
\def\sd{\relax{\rm \d\kern-0.5em /}}
\def\sk{\relax{\rm k\kern-0.5em /}}
\def\sn{\relax{\rm n\kern-0.5em /}}
\def\sl{\relax{\rm l\kern-0.5em /}}
\def\sP{\relax{\rm P\kern-0.7em /}}
\def\sBethe{\relax{\rm \Bethe\kern-0.5em /}}

\usepackage{varioref}
\usepackage{makeidx}
\makeindex

\newcommand\blfootnote[1]{%
  \begingroup
  \renewcommand\thefootnote{}\footnote{\hspace{-6mm}#1}%
  \addtocounter{footnote}{-1}%
  \endgroup
}

\usepackage[english]{babel}
\begin{document}


\thispagestyle{empty}

\renewcommand{\thefootnote}{\fnsymbol{footnote}}
\setcounter{page}{1}
\setcounter{footnote}{0}
\setcounter{figure}{0}

\vspace{-0.4in}

\begin{center}
$$$$
{\Large\textbf{\mathversion{bold}
The  Loom for General Fishnet CFTs
}\par}
\vspace{1.0cm}

\textrm{Vladimir Kazakov$^\text{\tiny a}$, Enrico Olivucci$^\text{\tiny b}$}
\blfootnote{\tt  \#@gmail.com\&/@\{e.olivucci\}}
\\ \vspace{1.2cm}
\footnotesize{
\textit{${}^\text{\tiny a}$ Laboratoire de Physique de l'\'{E}cole Normale Sup\'{e}ri\'{e}ure, CNRS, Universit\'{e} PSL, Sorbonne Universit\'{e}, Universit\'{e} Paris Cit\'{e}, 24 rue Lhomond, 75005 Paris, France}
\\
\textit{
${}^\text{\tiny b}$Perimeter Institute for Theoretical Physics,
Waterloo, Ontario N2L 2Y5, Canada}
\vspace{4mm}
}
\end{center}

\par\vspace{1.5cm}
\vspace{2mm}
\begin{abstract}

We propose a broad class of \(d\)-dimensional conformal field theories of $SU(N)$ adjoint scalar fields generalising the 4$d$ Fishnet CFT (FCFT) discovered by \"O.~G\"urdogan and one of the authors as a special limit of \(\gamma\)-deformed \(\mathcal{N}=4\) SYM~theory. In the planar
\(N\to\infty\) limit the FCFTs are
      dominated by the  ``fishnet" planar Feynman graphs. These graphs are explicitly integrable, as was shown long ago by  A.~Zamolodchikov. The Zamolodchikov's construction, based on the dual Baxter lattice (straight lines on the plane intersecting at arbitrary slopes) and the star-triangle identities, can serve as a ``loom" for ``weaving" the Feynman graphs of these  FCFTs,   with certain types of propagators, at any \(d\). The Baxter lattice with $M$ different slopes and any number of lines parallel to those, generates an FCFT consisting of \(M(M-1)\) fields and a certain number of chiral vertices of different valences with distinguished couplings. These non-unitary, logarithmic CFTs enjoy certain reality properties for their spectrum due to a symmetry similar to the PT-invariance of non-hermitian hamiltonians proposed by C.~Bender and  S.~Boettcher. We discuss in more detail the theories generated by a loom with $M=2,3,4$, and the generalisation of the loom FCFTs for spinning fields in 4d.
\end{abstract}

\newpage

\setcounter{page}{1}
\renewcommand{\thefootnote}{\arabic{footnote}}
\setcounter{footnote}{0}



{
\tableofcontents
}



\newpage
\section{Introduction}

The Fishnet Conformal Field Theories (FCFTs) first appeared in \cite{Gurdogan:2015csr} as a special double-scaling limit of the superconformal \(4d\)  \(\mathcal{N}=4\) SYM~theory, combining weak coupling and strong imaginary \(\gamma\)-deformation. Its particular bi-scalar case appeared to be explicitly integrable in the planar, multicolor limit due to the observation by A.~Zamolodchikov ~\cite{Zamolodchikov:1980mb} that the ``fishnet" planar Feynman graphs with the shape of a regular square lattice, dominating in the perturbation theory, are integrable.~\footnote{This contrasts to the still mysterious planar integrability of the full \(\mathcal{N}=4\) SYM~theory, as well as of its general fishnet limit  with three couplings~\cite{Gurdogan:2015csr}  dominated by so called dynamical fishnet graphs~\cite{Kazakov:2018gcy}, still awaiting  explanation.} Since then, a few generalizations of this FCFT have been found, for various dimensions and massless propagators~\cite{Kazakov:2018qbr}, inclusion of fermions~\cite{Kazakov:2018gcy} as well as the fishnet limit of 3-dimensional \(\gamma\)-deformed ABJM theory~\cite{Caetano:2016ydc}.  

The FCFTs are genuinely non-unitary, logarithmic CFTs. The study of their physical properties in the planar limit can be achieved in much greater detail due to the narrow set of planar Feynman graphs in the theory, in addition to their integrability. A particularly studied case is the \(d\)-dimensional bi-scalar CFT. Certain sets of conformal dimensions of its local operators  have been computed explicitly, using only conformal symmetry for the shortest ones~\cite{Grabner:2017pgm,Gromov:2018hut,Kazakov:2018qbr}. The conformal dimensions of certain longer operators have been calculated using the equivalence to the integrable non-compact spin chain, via the Quantum Spectral Curve approach~\cite{Gromov:2013pga,Gromov:2014caa,Kazakov:2015efa,Kazakov:2018ugh},~perturbatively up to very high orders and numerically with a great precision~\cite{Gromov:2017cja}, or in the asymptotic limit of long operators~\cite{Caetano:2016ydc}. The thermodynamical Bethe ansatz equations have been written for general types of such operators~\cite{Basso:2018agi,Basso:2019xay}. Certain structure constants have been computed,  exactly~\cite{Grabner:2017pgm,Gromov:2018hut,Kazakov:2018qbr} or in various approximations~\cite{Basso:2015zoa}. Four-point correlators of certain operators have been computed exactly,  in disc  topology~\cite{Basso:2017jwq,Derkachov:2018rot,Derkachov:2019tzo, Derkachov:2020zvv,Derkachov:2021ufp} and for short operators  in cylindrical topology \cite{Grabner:2017pgm,Gromov:2018hut,Kazakov:2018qbr}.  It was shown that the bi-scalar  FCFT possesses the quantum mechanically stable flat vacua~\cite{Karananas:2019fox}. The fishnet amplitudes appear to be dominated by a single fishnet graph and obey the Yangian symmetry~\cite{Chicherin:2017cns,Chicherin:2017frs, Corcoran:2021gda, Duhr:2022pch}. Finally, the AdS dual of the FCFT has been proposed in the form of a  ``fish-chain" -  a discretised string theory living on AdS\(_5\) background \cite{Gromov:2019aku, Gromov:2019bsj, Gromov:2019jfh}. 

However, only a limited subset of all possible FCFTs following from the Zamolodchikov's construction of integrable fishnet graphs has been explored by now. In this work we will fully exploit this construction and show how to define the most general FCFTs dominated in the planar limit by integrable fishnet graphs, dual to the Baxter lattice of an arbitrary number \(n\) of  intersecting straight lines having \(M\le n\) different slopes (so the rest of them are parallel to the basic \(M\) ones). We will show that such FCFTs have \(M(M-1)\) scalar complex \(N\times N\) matrix fields with a certain number of interactions (with free couplings) whose number depends on \(M\).
We will call such a Baxter lattice a Loom -- the device to ``weave"  dual graphs, and the corresponding field theories -- the Loom FCFTs.
They will be classified and counted in this paper. 

At each order of perturbation theory of such a Loom FCFT we have to sum up all Feynman graphs dual to the Baxter lattices with all possible types of intersections of these \(n\) lines.  We will also discuss the renormalisation and clarify the issue of the multi-trace couplings~\cite{Sieg:2016vap} which  have to be added to the action in order to preserve the conformal symmetry in such generalised FCFTs. Furthermore we will consider a few new concrete realisations of the Loom FCFTs. We give the detailed description of the cases
of \(M=3 \)  and \(M=4\) slopes.

An important property of these non-unitary FCFTs is a symmetry of their Lagrangians w.r.t. the time-reversal, \(T\)-transformation --  i.e. the simultaneous hermitian conjugation of all matrix fields -- and the \(t\)-transformation -- i.e. the transposition of all matrix fields. We called it the ``\(tT\)-symmetry". This leads to certain positivity properties of the spectrum of anomalous dimensions: the dimensions can be either real or in complex conjugate pairs. This property is reminiscent of the spectrum of non-hermitian \(PT\) symmetric QM hamiltonians proposed in~\cite{Bender:1998ke,Bender:2007nj}.

\section{Integrable planar Feynman graphs}
\label{sec:graphs}

The Loom FCFTs we are going to construct in this paper will be called integrable in the planar limit, meaning that the planar Feynman diagrams defining various physical quantities in such FCFTs are integrable at any loop order. The integrability of a Feynman diagram should be understood in the sense of A.~Zamolodchikov's construction \cite{Zamolodchikov:1980mb} of integrable \(2d\) statistical mechanical systems that imitate these graphs. In this section, we will review the Zamolodchikov's construction. It is based on the general Baxter lattice -- a set of intersecting straight lines on the plane. The Baxter lattices are in one-to-one correspondence with the  Feynman diagrams of the theory, whose construction will be reviewed below.  

In \cite{Zamolodchikov:1980mb} it was showed that Feynman diagrams with triangular, square and hexagonal lattice topology are integrable.
The FCFTs dominated by such graphs have been constructed in~\cite{Gurdogan:2015csr,Caetano:2016ydc,Mamroud:2017uyz}.~ But the Loom FCFTs we will present in this paper are based on the most general integrable graphs of Zamolodchikov's construction. The integrability is verified in all these cases via Baxter's \emph{star-triangle} relation \cite{Baxter:1972hz, DEramo:1971hnd}.

Let us review now the construction of the Feynman diagrams dual to the Baxter lattices.

\subsection{Feynman diagrams vs dual graphs}

First, we start from an arbitrary scalar planar diagram with dimensionless vertices and define its dual graph. We will show that the integrability of the diagram implies that the dual graph is of the Baxter type, i.e. made of straight lines. 

The central object of our construction are Feynman diagrams in $d$ dimensions with massless propagators. In order to define planarity and the \(1/N \) expansion over the topologies of diagrams, we work with the fields in certain matrix representation -- we choose the $SU(N$) adjoint representation -- so to impose a rigid order for the fields inside each vertex. Each propagator of such a field $\Phi_{ij}$ scales with a given dimension $\Delta$ and has the matrix structure
\begin{equation}
\label{propagator}
G_{\Delta}(x)_{ij,kl} = \frac{\delta_{ik} \delta_{jl}-N^{-1}\delta_{ij}\delta_{kl}}{x^{2\Delta}}\,.
\end{equation}
 Each  Feynman graph consisting of propagators of this type has, after contraction of all indices,  a standard weight \(N^{2-2h}\), where \(h\)  is the genus of the graph~\cite{tHooft:1973alw}. The propagators can be convoluted at their endpoints forming vertices of a given valence $n\geq 3$, that is
\begin{equation}
 \int d^d z \prod_{k=1}^n G_{\Delta_k}(x_k-z)\,,
\end{equation}
where $G_{\Delta}(x-y) =(1/(x-y)^2)^{\Delta}$ is a propagator already stripped of its indices. For  the scale invariance (and thus the conformal invariance) each vertex is subject to the constraint $\sum_{k=1}^n \Delta_k= d$ in order for the associated coupling constant to be dimensionless.~\footnote{We will use for the actions of these FCFTs the normalization \(S=N\text{Tr}  (\dots)\), so that all these couplings are finite in the large \(N \)  limit ('t~Hooft couplings).} 
\begin{figure}[h]
\begin{center}
\includegraphics[scale=1.0]{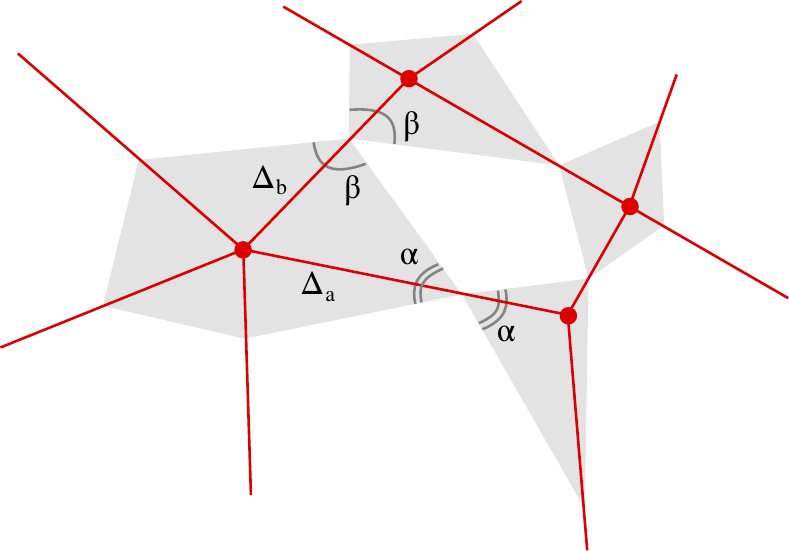}
\end{center}
\caption{A Feynman diagram with four scale-invariant vertices and its associated dual graph with the "checkerboard" coloring. The scaling dimensions $\Delta_a$ and $\Delta_b$ are related to the angles $\alpha$ and $\beta$ respectively. }
\label{dual_construction}
\end{figure}
\begin{figure}[h]
\begin{center}
\includegraphics[scale=1.2]{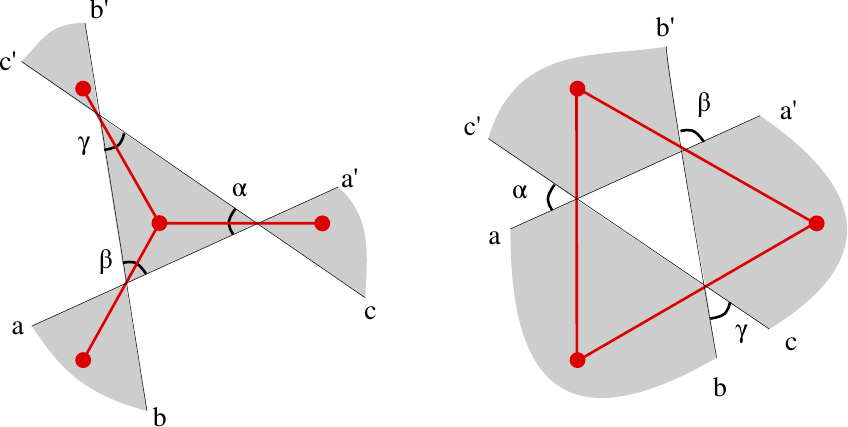}
\caption{Star-triangle duality. The faces of the dual graph are black/white. Red lines and dots correspond to the Feynman diagram elements. The graph on the right is obtained by that on the left moving the line \([a,a']\) across the intersection of \([b,b']\) and \([c,c']\) lines.}
\label{STR2}
\end{center}
\end{figure}
\begin{figure}[h]
\begin{center}
\includegraphics[scale=0.8]{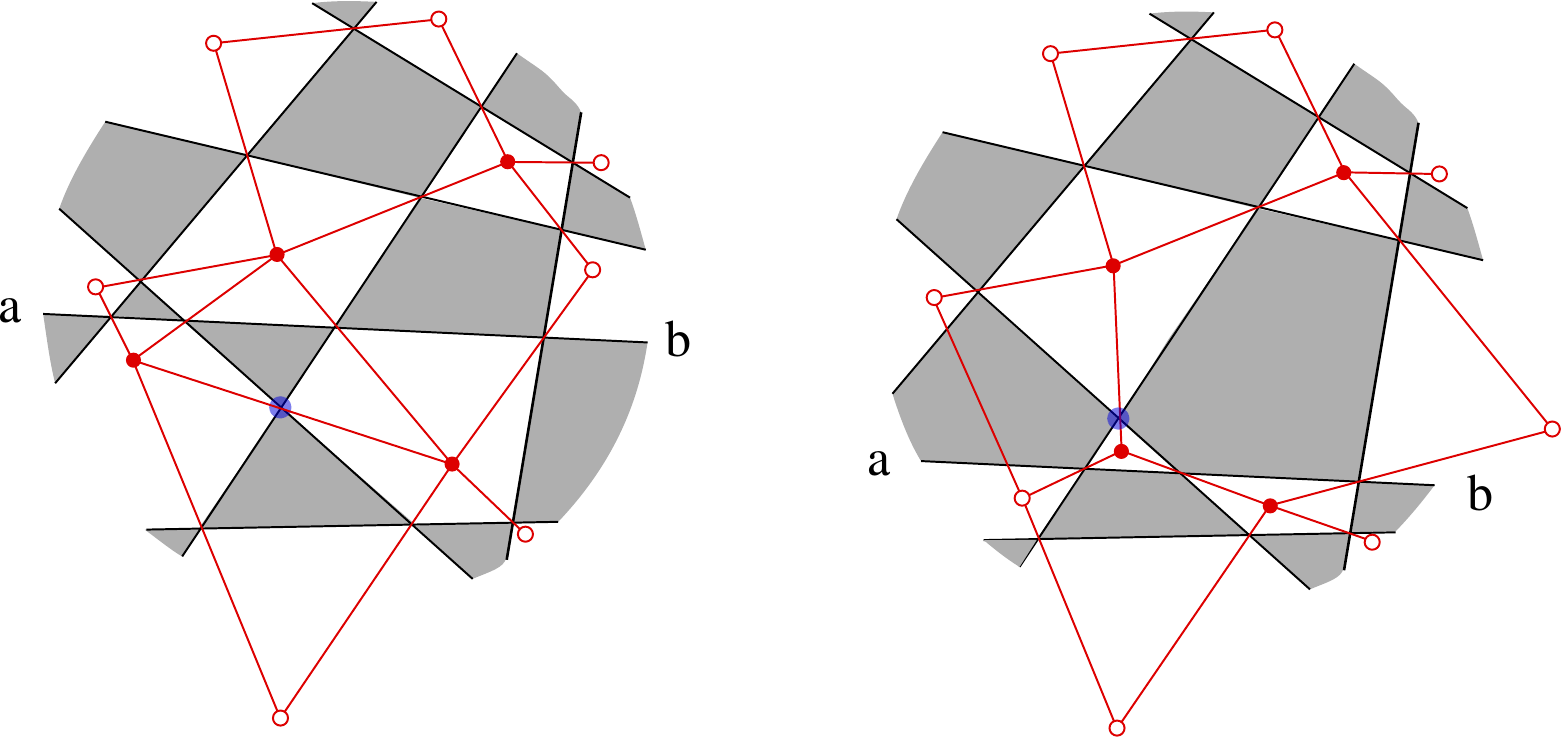}
\end{center}
\caption{A dual graph made of coloured faces and its dual Feynman diagram, made of propagators (red lines), integrated points (red dots) and external points (red circles). The graph on the right is obtained from graph on the left via the translation of the segment \([a,b]\) across the vertex marked in blue. The two diagrams are equal by star-triangle identity.}
\label{loom_circle}
\end{figure}
\begin{figure}[h]
\begin{center}
\includegraphics[scale=1.1]{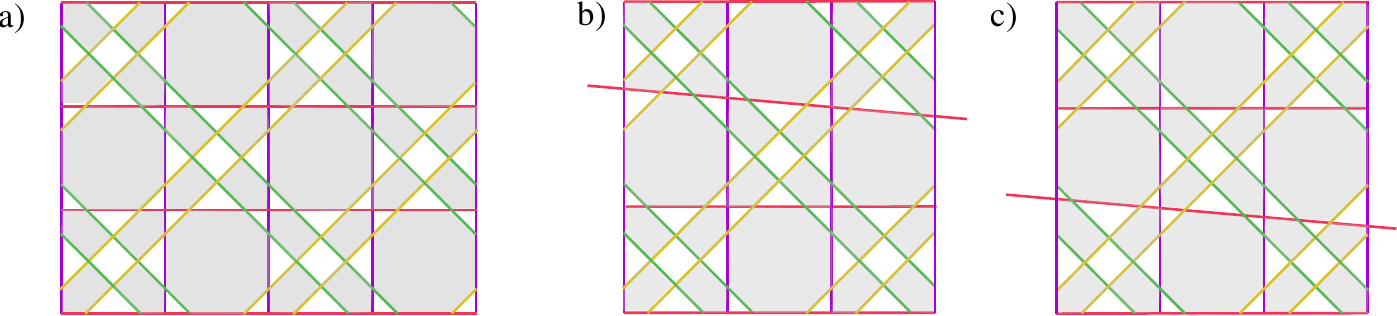}
\caption{A portion of Baxter lattice with $4$ slopes depicted by different colours (a). The same graph after tilting one red line (b). On the associated Feynman diagram the tilting corresponds to twisting the powers of propagators across the red line. The graph enjoys the symmetry (b)$\sim$(c) of the loom via the star-triangle relation. On the Feynman diagram this amounts to the commutation of an infinite family of integral kernels -- the portion of Feynman diagram associated to the tilted line and parametrised by the tilting angle -- with the rest of the diagram. This fact establishes the integrability of the diagrams.}
\label{integr}
\end{center}
\end{figure}

The \emph{dual graph} of a Feynman diagram is defined by the following prescription \cite{Zamolodchikov:1980mb} (see Fig.~\ref{dual_construction}):
\begin{itemize}
\item A vertex of the Feynman diagram with $n$ propagators lies inside a polygon whose $n$ edges belong to the dual graph and each of its vertices is traversed by  a propagator originating in the middle of the  polygon.
\item Each angle \(\alpha\) of the $n$ internal angles of the polygon is determined by the scaling dimension \(\Delta\) of the propagator that passes through this angle, according to the formula
\begin{equation}
\label{angle_delta}
\pi-\alpha = \frac{2\pi}{d} \Delta\,.
\end{equation}
\item The faces of the dual graph feature a checkerboard colouring (black/white), depending on whether they contain or not a Feynman diagram vertex.
\end{itemize}
\begin{figure}[h]
\begin{center}
\includegraphics[scale=1.2]{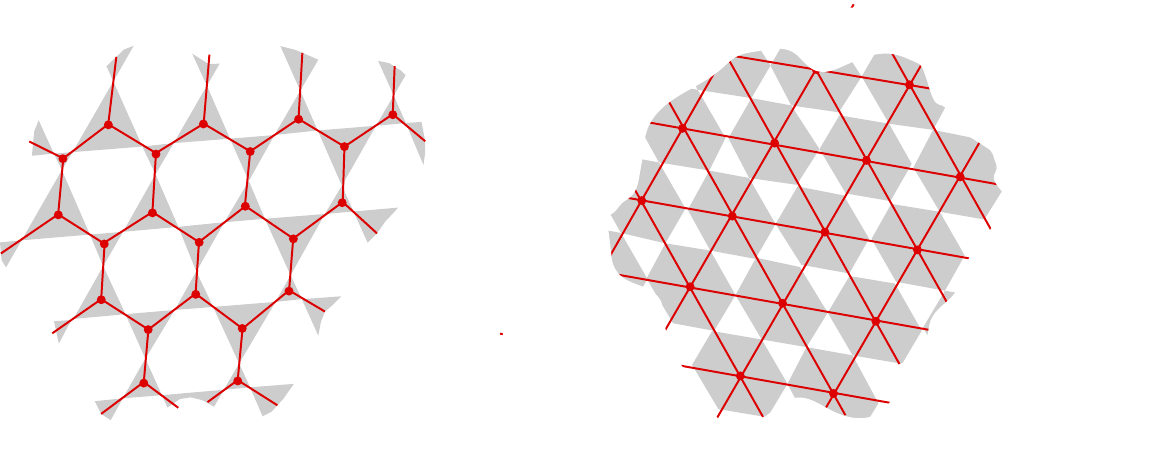}
\caption{a) Hexagonal ``honeycomb" lattice Feynman graph (red propagators and vertices) the dual graph with Kagom\'{e} topology. b) Triangular lattice Feynman diagram. The Baxter lattice is the same Kagom\'{e} as in (a), with exchanged colouring of the faces.}
\label{honey1}
\vspace{2mm}
\includegraphics[scale=1.2]{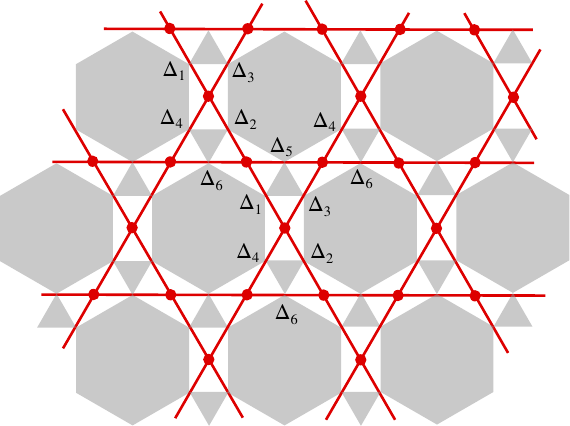}
\caption{Feynman diagram with Kagom\'{e} topology, in red. The dual Baxter lattice is a tiling of the plane with black/white faces that cannot be realised with straight lines.}
\label{3scalars}
\end{center}
\end{figure}
Notice that the sum of the interior angles of each $n$-gon is equal to $(n-2)\pi$, which encodes the scale-invariance condition $\sum_{i \in \text{vertex}} \Delta_i = d $ of the vertices of the diagram via \eqref{angle_delta}.
Conversely, for a given dual graph one can draw the original Feynman diagram by inserting a vertex inside each black (either white) face and connecting by a propagator any pair of vertices of the diagram that lie on faces of the dual graph that have a vertex in common. The power (scaling dimension of the corresponding field) of the propagator is determined by the angle of the face of the dual graph through which it passes.

\subsection{Loom for integrable Feynman graphs}

The geometry of the dual graph is useful to identify the restricted subclass of all conformal diagrams that are integrable. The precise criterion for identifying the integrable graphs will be given below. Tentatively, \emph{integrability} means that the graphs can be computed via the methods of quantum integrability, following the Zamolodchikov's construction~\cite{Zamolodchikov:1980mb} and using the tools of non-compact integrable spin-chains with conformal \(SO(1,d+1)\) symmetry~\cite{Chicherin:2013rma,Chicherin:2012yn, Derkachov:2021rrf}  developed on the basis of~\cite{Zamolodchikov:1980mb}.
It will also help us to write the Lagrangians of FCFTs that are dominated only by the set of these integrable planar diagrams (or some of their subsets). 
We can relate the integrability of a diagram to the geometry of its dual graph. The main statement on the basis of our construction is:

\textit{ A Feynman diagram is integrable if its dual graph is made of intersecting straight lines, i.e. the Feynman diagram is dual to a Baxter lattice. } 

The proof of this statement relies on the star-triangle identity which is the fundamental equation of integrability of the $SO(1,1+d)$ spin chain. The star-triangle identity for both, a fragment of  Baxter graph and  the dual Feynman diagram, is presented on the Fig.~\ref{STR2}. On the Feynman diagram side the star-triangle is an integral identity that equates a cubic scale-invariant vertex to the product of three propagators~\cite{DEramo:1971hnd}~(see figure~\ref{STR2}):
\begin{equation}
\int d^d z \prod_{k=1}^3 G_{\Delta_k}(x_k-z) = \pi^{\frac{d}{2}} \frac{\Gamma\left(\frac{d}{2}-\Delta_1 \right)\Gamma\left(\frac{d}{2}-\Delta_2 \right)\Gamma\left(\frac{d}{2}-\Delta_3 \right)}{\Gamma\left(\Delta_1\right)\Gamma\left(\Delta_2\right)\Gamma\left(\Delta_3\right)} \prod_{k=1}^3 G_{\frac{d}{2}-{\Delta_k}}(x_{k-1}-x_{k+1}) \,.
\label{STR}
\end{equation}

If the dual graph is made of straight lines, the four segments around a graph vertex have pairwise the same angles. Hence, each  vertex of the graph features four angles with  equal pairs of opposite angles, two angles on the black faces and two supplementary on the white faces. Consequently,  one can draw two conformal Feynman diagrams for the same dual Baxter graph: one has vertices inside black faces, while the other has vertices inside white faces. The scaling dimensions of the  propagators of  two  diagrams crossing the same   vertex are   related as
\begin{equation}
\Delta'  = \frac{d}{2}-\Delta=\frac{d}{2\pi}\alpha \,.
\end{equation}
Let us stress that the   propagators with both dimensions can be present in each of these diagrams if we admit the presence  of parallel lines in the loom.  

The sharp statement of integrability can be seen from a Feynman integral once we move a line (conserving its slope) through  one or a few crossings  in the loom, as depicted in figures \ref{loom_circle},\ref{integr}. Such moves can be represented as commutations of certain  convoluted integral kernels in the associated pieces of the corresponding Feynman diagram. The exchange of lines is therefore a statement of commutation of integral operators -- certain monodromy matrices of the inhomogeneous  quantum chain  with the symmetry $SO(1,d+1),$ with conformal spins represented by  the spacetime coordinates of the vertices. 

\subsection{Integrable vs non-integrable diagrams: examples}

Let us consider a few examples of planar conformal diagrams, and determine whether or not they belong to the loom picture, i.e. whether they are \emph{integrable}.

First, we take a diagram with the topology of a honeycomb, formed by cubic vertices of two types (see Fig.\ref{honey1} (left)). The scaling dimensions of the three propagators  converging in each vertex are $\Delta_1$, $\Delta_2$ and $\Delta_3$. If we impose the scale-invariance constraint $\Delta_1+\Delta_2+\Delta_3=d$ the dual graph has the topology of a Kagom\'{e} lattice formed by straight lines with three different slopes. Hence, the integrability definition is satisfied.

Similarly, let us take a diagram with the topology of a triangular lattice made of valence-$6$ vertices (see Fig.\ref{honey1} (right)). The dual graph has the topology of a Kagom\'{e} -- and so it is made of straight lines -- if and only if the six propagators emitted by a vertex and ordered clockwise, have dimensions $\Delta_1=\Delta_4$, $\Delta_2=\Delta_5$ and $\Delta_3=\Delta_6$ and also satisfy the scale-invariance condition $\Delta_1+\Delta_2+\Delta_3=d/2$. The checkerboard colouring of faces is interchanged with respect to the previous example.
Thus, also this Feynman graph is integrable.

Let's analyse the graphs generated starting from the Kagom\'{e} Baxter lattice, and the associated Feynman diagrams. By moves of lines we can relate the honeycomb and triangular lattice examples, that is to exchange the checkerboard colouring of the faces. Furthermore, we could form squares and pentagons in the loom, besides hexagonal and triangular faces, which means to generate quartic and quintic interaction vertices in the Feynman diagrams. We will use these moves in the next section to construct all interactions in the related Loom FCFTs.

It is interesting to consider now those Feynman diagrams with the  Kagom\'{e} structure, i.e. made up of three types of quartic vertices and propagators depicted in Fig.\ref{3scalars}. These diagrams have an important role in the dynamical Fishnet models  and in the Eclectic field theory \cite{Kazakov:2018gcy,Ahn:2020zly}. The dimensions of the four propagators for each vertex can be labelled as $(\Delta_1,\Delta_2,\Delta_3,\Delta_4)$,  $(\Delta_1,\Delta_2,\Delta_5,\Delta_6)$ and  $(\Delta_3,\Delta_4,\Delta_5,\Delta_6)$. The dual graph of such a Feynman diagram is a tiling made of white rectangles and black triangles and hexagons which cannot be realised with straight lines (see Fig.\ref{3scalars}). Indeed, in terms of scaling dimensions that would require the following system of linear relations 
\begin{align}
\begin{aligned}
&\Delta_1+\Delta_2=\Delta_3+\Delta_4=\Delta_5+\Delta_6=d/4\,,\\
&\Delta_1+\Delta_3+\Delta_5=d/2\,,\\
&\Delta_2+\Delta_4+\Delta_6=d/2\,.
\end{aligned}
\end{align}
 which has no solution. We conclude that a planar conformal diagram made of these three quartic vertices cannot belong to a Loom and thus it is not integrable (at least by the star-triangle method). 

In what follows, using the observations of this section we will construct the most general FCFTs whose planar diagrams  are integrable, i.e. each of them dual to its Loom --  a particular  Baxter graph  consisting of a certain number \(n\) of lines with \(M\le n\)  different slopes.

\section{FCFTs from the Loom}
\label{sec:LoomFCFT}

In this chapter, we will formulate the Lagrangian of a general Loom FCFT, integrable in the planar limit. We will start from the case of FCFTs with \(M=2,3,4\) slopes of lines on the Baxter lattice, and later we will extend our scope to the general \(M\)-Loom FCFTs.

The Loom FCFTs can be called  ``solvable", in the sense that many interesting physical quantities can be efficiently computed at high orders of perturbation theory, or even at finite coupling, using the integrability property of the underlying Feynman diagrams. 
 
In a general, $M$-Loom FCFT one  introduces a set of   \(M(M-1) \) complex matrix quantum fields with particular dimensions and interaction vertices. The number of such interactions, which we will describe and count below, is quickly growing  with \(M\). The realisation of a Loom FCFT requires to restrict the possible vertices to a fixed chirality, so that to get rid of all the diagrams that cannot be generated via the Baxter lattices. The chirality of a vertex is defined by the order in which the complex matrix fields -- in some representation of $SU(N)$ -- appear inside a single trace interaction. 
It is clear then that choosing specific chirality -- as one always does in the Loom FCFT -- requires to deal with non-unitary, logarithmic CFT.  Nevertheless, as we will show in section~\ref{sec:tTsym}, these FCFTs enjoy an interesting symmetry, similar to the \(PT\) invariance of non-hermitian QM hamiltonians proposed in~\cite{Bender:1998ke,Bender:2007nj}, which leads to certain reality properties for the spectrum of conformal dimensions of the operators.

We will also discuss the renormalization of the Loom FCFTs and the necessity of adding various multi-trace terms into the Lagrangian, necessary for restoring the conformal invariance.

\subsection{Bi-scalar fishnet}
The simplest example of the Loom FCFT is given by the bi-scalar fishnet \cite{Gurdogan:2015csr,Kazakov:2018gcy}. This is the case of a loom with $M=2$ slopes, i.e. there are  $M(M-1)=2$ different angles on the Baxter lattice. Hence the Feynman diagrams have two types of propagators \eqref{propagator} associated with two fields $X$ and $Y$. The only possible type of graphs  is then the square lattice with vertices the $\text{Tr}\left[XY \bar X \bar Y \right]$. The angles between  two loom directions are $\theta_{12}$ and $\pi-\theta_{12}$ and for the isotropic Fishnet one has $\theta_{12}=\pi/2$ -- that is the lattice of two  orthogonal sets of lines. 

The Lagrangian of bi-scalar FCFT is given  by
\begin{equation}\label{FCFTbiscalar}
    {\cal L}_{d}^{(a)}= \! N\,\Tr
    \left[   \bar X \, (-\p_\mu \p^\mu)^{d/2-\Delta_1}\, X
+\bar Y \, (-\p_\mu \p^\mu)^{d/2-\Delta_2}\, Y+ (4 \pi)^{d/2} \xi^2\,\, XY \bar X \bar Y\right]\!+\text{double-traces},
  \end{equation}
where $X,Y$ are complex scalar fields and $\Delta_2=d/2-\Delta_1 = d \,\theta_{12}/(2\pi)$.
The theory is also invariant under the internal symmetry $U(1)\otimes U(1)$
\begin{equation}
X \to e^{i a} X\,,\,\, Y \to e^{i b} Y\,,
\end{equation}
since the quartic vertex has zero charge. 
 The double-trace terms here, needed for the renormalisation of the theory and for fixing eventually the corresponding couplings at the conformal point, are 
\begin{equation}
\text{Tr}\left[XY \right] \text{Tr}\left[\bar X \bar Y\right]\,,\, \text{Tr}\left[\bar XY\right]  \text{Tr}\left[\bar Y X\right]\,.
\label{XYdoubletraces}
\end{equation}
The bi-scalar theory at $\theta_{12}=\pi/2$ has extra internal symmetries
\begin{equation}
X \to Y\,, \, Y\to \bar X\quad \text{or}\,\,\,\,\,X \to \bar Y\,, \, Y\to X\,,
\label{XXYYdoubletraces}
\end{equation}
At this special point, further double-trace counter-terms are needed for the UV completeness of the theory, namely
\begin{equation}
\label{XXbXbX}
\text{Tr}\left[X^2 \right] \text{Tr}\left[\bar X ^2\right]\,,\, \text{Tr}\left[Y^2\right]  \text{Tr}\left[\bar Y ^2\right]\,,
\end{equation}
which would otherwise be dimensionful. All couplings for double-trace counter-terms in this theory have real or complex conjugate critical points as  functions of the (non-renormalised) coupling $\xi$.  The conformal symmetry is restored at such critical points~ \cite{Sieg:2016vap,Grabner:2017pgm,Kazakov:2018gcy}.

\subsection{Loom FCFT with 3 slopes}

Let us consider now the Loom FCFT for $M=3$ slopes in $d$-dimensional Euclidean spacetime.  In this case the field content of the theory will be that of $M(M-1)=6$ complex scalars in the adjoint representation of $SU(N)$. We denote  three of these fields as $X,Y,Z$ and $\Delta_1,\Delta_2,\Delta_3$ are their scaling dimensions. These three fields interact among themselves via a sextic vertex, and satisfy the scale-invariance constraint $\Delta_1+\Delta_2+\Delta_3=d/2$. If they were alone, they would form the planar Feynman diagrams of the shape of  triangular lattice (at least in the bulk of a large diagram), similarly to the Fishnet reduction of ABJM theory~\cite{Caetano:2016ydc}. The dual Baxter lattice corresponding to such diagrams forms the Kagom\'{e} lattice with hexagonal and triangular faces (see the right figure~\ref{honey1}).
\begin{figure}[h]
\begin{center}
\includegraphics[scale=1.4]{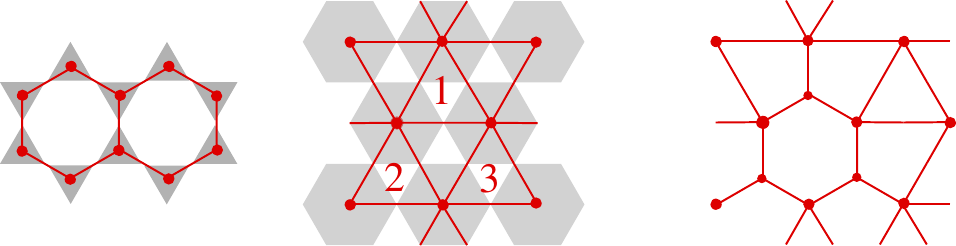}
\end{center}
\caption{Feynman diagram structures generated by a Baxter, loom lattice with $3$ slopes. First two examples are the hexagonal honeycomb and the triangular lattice. The third picture is a rather generic loom structure involving also vertices of valence $4$ and $5$, obtained from the second picture, applying star-triangle to the three triangles marked by numbers.}
\end{figure}

Similarly, we denote the other three fields  as $u,v,w$ and notice that they interact via the cubic vertices, giving rise to diagrams with honeycomb structure  in the bulk, as in  the left figure~\ref{honey1}. The corresponding Baxter lattice is once again of Kagom\'{e} type, but now the diagram vertices lie on triangular faces rather than on hexagonal ones. The dimensions of  $u,v,w$-fields are $d/2-\Delta_1,d/2-\Delta_2,d/2-\Delta_3$, respectively. The fields $u,v,w$ can be considered as \emph{dual} to the fields $X,Y,Z$, in the sense that their propagators cross respectively the same vertices of the Baxter lattice but across the complementary pair of angles.

\begin{figure}
\includegraphics[scale=0.63]{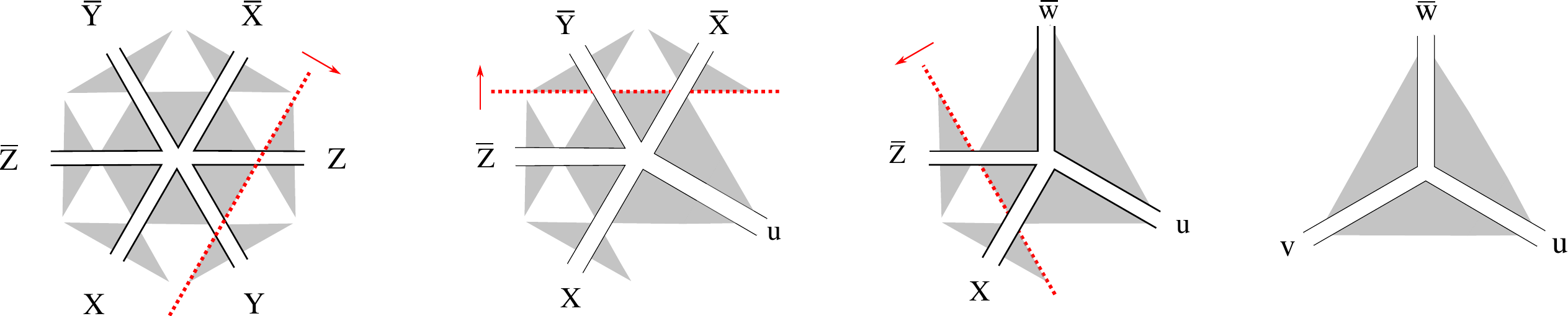}
\caption{A primer of the single-trace interaction in double-line notation. The vertex of valence six on the left can be used to generate the vertices of lower valence by substituting two propagators of fields $X,Y,Z$ with one propagator of a corresponding field $u,v,w$, \eqref{replacement}. On the Baxter lattice this consists of  moving out the red dotted line at each step.}
\label{3Loom_vertices}
\end{figure}

The analysis of the interaction vertices is demonstrated on Fig.~\ref{3Loom_vertices}. All of them can be generated starting from the Kagom\'{e} graph with vertices inside the hexagonal faces, corresponding to a Feynman diagram with the topology of a triangular lattice, with a sextic interaction that reads
\begin{equation}
\text{Tr}\left[X Y Z \bar X \bar Y \bar Z \right]\, ,
\label{sextic}
\end{equation}
It is depicted on the left picture of Fig.~\ref{3Loom_vertices}.
On the dual lattice, this vertex is formed by three pairs of straight lines, one pair for each of three slopes.  

By star-triangle moves of loom lines, the hexagonal faces that contain the valence-$6$ vertex can be transformed into faces of lower number of edges, that is to vertices of a lower valence. 
The replacement of two among $X,Y,Z$ with one of the  fields $u,v,w$ follows the pattern~\footnote{Notice that the definition of \(\bar w,\bar u, \bar v\) does not follow the rules of hermitian conjugation. As we already mentioned, in these non-unitary theories we have to consider in the functional integral all the barred fields as independent from non-barred.}
\begin{equation}
\label{replacement}
X Y \mapsto w\,,\,\,YZ \mapsto u\,,\,\, Z \bX \mapsto v\,,\,\, \bar X\bar  Y \mapsto \bar w\,,\,\, \bar Y\bar Z \mapsto \bar u\,, \bar ZX \mapsto \bar v\,.
\end{equation}
Any of this moves transforms the sextic vertex  into one of the six quintic vertices
\begin{align}
\begin{aligned}
&\text{Tr}\left[X Y v \bY \bZ \right],\,\text{Tr}\left[w Z \bX \bY \bZ \right],\,\text{Tr}\left[X u \bX \bY \bZ \right]\,,\\
&\text{Tr}\left[X Y Z \bX \bar u \right],\,\text{Tr}\left[X Y Z \bar w  \bZ \right],\,\text{Tr}\left[\bar v Y Z \bX \bY  \right]\,,
\end{aligned}
\label{quintic}
\end{align}
as it is demonstrated on the 2nd picture on Fig.~\ref{3Loom_vertices}.
One further replacement of fields delivers nine valence-$4$ vertices
\begin{align}
\begin{aligned}
&\text{Tr}\left[w v \bY \bZ \right],\,\text{Tr}\left[w Z \bX \bar u \right],\,\text{Tr}\left[X u \bar w \bZ \right]\,,\\
&\text{Tr}\left[X u \bX \bar u \right],\,\text{Tr}\left[\bar v Y Z \bar w   \right],\,\text{Tr}\left[\bar v Y v \bY  \right]\,,\\
&\text{Tr}\left[X Y v \bar u \right],\,\text{Tr}\left[w Z \bar w \bZ \right],\,\text{Tr}\left[\bar v u \bX \bY \right]\,,
\end{aligned}
\label{quartic}
\end{align}
as demonstrated on the 3nd picture on Fig.~\ref{3Loom_vertices}. And finally, the last move on the loom transforms the initial sextic vertex to the same graph with interchanged white/black faces, corresponding to a ``honeycomb" Feynman diagram with cubic vertices, as seen in the left picture of   Fig.~\ref{3Loom_vertices}:
\begin{equation}
\text{Tr}\left[ \bar u w v\right]\,,\,\,\, \text{Tr}\left[u \bar w \bar v \right]\,.
\label{cubic}
\end{equation}

All the $18$ vertices of the theory are neither hermitian nor are they accompanied by their hermitian conjugate vertices  in the Lagrangian. That is, all single-trace interactions have only one chirality, and the theory is not unitary. Each interaction vertex carries an independent coupling constant. We will usually assume that the couplings associated to a pair of vertices related by the exchange of barred/unbarred fields are complex conjugate of each other. This assumption, although not necessary for a well defined integrable FCFT, enables the Loom FCFT with some reality property of the spectrum, explained later in \ref{sec:tTsym}.

 The Lagrangian of $M=3$ Loom FCFT reads
\begin{align}
\begin{aligned}
\label{3loom_L}
\mathcal{L}_{3-loom}  &=N\text{Tr}\left[\bar X(-\partial_{\mu} \partial^{\mu})^{\frac{d}{2}-\Delta_1} X +\bar Y(-\partial_{\mu} \partial^{\mu})^{\frac{d}{2}-\Delta_2} Y +\bar Z(-\partial_{\mu} \partial^{\mu})^{\frac{d}{2}-\Delta_3} Z \right] + \\&+N\text{Tr}\left[\bar u(-\partial_{\mu} \partial^{\mu})^{\Delta_1} u +\bar v(-\partial_{\mu} \partial^{\mu})^{\Delta_2} v +\bar w(-\partial_{\mu} \partial^{\mu})^{\Delta_3} w \right] +\\&+\mathcal{L}_{\text{sT}}  + \mathcal{L}_{\text{dT}} \,,
\end{aligned}
\end{align}
where the single-trace part $ \mathcal{L}_{\text{sT}}$ of the Lagrangian is a sum over the $18$ vertices listed in~\eqref{sextic},\eqref{quintic},\eqref{quartic},\eqref{cubic}, generically all with independent  coupling constants.  The part $\mathcal{L}_{dT}$ contains the double-trace counter-terms that need to be included in order to renormalise some composite operators and will be analysed in section~\ref{renom_3}.

The $3$-loom FCFT has an internal symmetry $U(1)\otimes U(1) \otimes U(1)$. Under this symmetry the fields transform acquiring a phase
\begin{equation}
\phi \,\, \to \,\, \phi \,e^{i(\alpha_1 q_1 +\alpha_2 q_2+ \alpha_3 q_3)}\,,
\end{equation}
that depends on the quantum numbers, \(U(1)\) charges $(q_1,q_2,q_3)$ listed in the following tables
 \label{tab_charge}
  \begin{center}
    \begin{tabular}{c|c|c|c|} 
       &$q_1$ & $q_2$ & $q_3$ \\
      \hline
      $X$ & 1 & 0 &0\\
      $Y$ & 0 & 1 & 0\\
      $Z$ & 0 & 0&1
             \end{tabular}\;\;\; ; \;\;\;
           \begin{tabular}{c|c|c|c|} 
       &$q_1$ & $q_2$ & $q_3$ \\
      \hline
       $u$ & 0 & 1 &1\\
      $v$ & -1 & 0 &1\\
      $w$ & 1 & 1&0
       \end{tabular}
  \end{center}
 The barred   fields have  opposite quantum numbers. The charges for $u,v,w$ are fixed by requiring all interaction vertices to be neutral, that is such that both sided of the replacements \eqref{replacement} have the same quantum numbers.
 
A notable reduction of the $3$-Loom FCFT is the single-coupling Lagrangian
\begin{equation}
\label{ABJM_II}
\mathcal{L}_{ABJM} \!=N\! \text{Tr}\left[\bar X(-\partial_{\mu} \partial^{\mu})^{\frac{d}{2}-\Delta_1} X +\bar Y(-\partial_{\mu} \partial^{\mu})^{\frac{d}{2}-\Delta_2} Y +\bar Z(-\partial_{\mu} \partial^{\mu})^{\frac{d}{2}-\Delta_3} Z  + \eta^2 XYZ\bar X \bar Y \bar Z \right]\,.
\end{equation}
 It  reduces to the ABJM fishnet CFT~\cite{Caetano:2016ydc} when $\Delta_1=\Delta_2=\Delta_3 =1/2;\,  d=3$ and in this case the $U(1)$ symmetry is the residual R-symmetry group after the  breaking of original superconformal symmetry of the ABJM gauge theory.  
 To be precise, let us point out that the ABJM FCFT should be defined with fields $X,Z$ in the fundamental and $Y$ in the anti-fundamental representation of $U(N)$. Practically, the Lagrangian \eqref{ABJM_II} produces the same correlators as in the properly defined ABJM fishnet.

Another interesting reduction of 3-loom FCFT contains only triple vertices:
\begin{equation}
\label{ABJM_III}
\mathcal{L}_{honeycomb} \!=N \,\text{Tr}\left[\bar u(-\partial_{\mu} \partial^{\mu})^{\Delta_1} u +\bar v(-\partial_{\mu} \partial^{\mu})^{\Delta_2} v +\bar w(-\partial_{\mu} \partial^{\mu})^{\Delta_3} w +\,\zeta_1^2  \bar u w v\,+\,\,\zeta_2 ^2\, u \bar w \bar v\right], 
\end{equation}
which gives at \(d=6,\,\Delta_1=\Delta_2=\Delta_3=1\)  the FCFT proposed in~\cite{Mamroud:2017uyz}.

Under special circumstances the 3-Loom theory \eqref{3loom_L} enjoys discrete symmetries, following from the fact that any cyclic re-labelling of loom directions -- and therefore of the fields -- does not change the set of single-trace interactions.
 Indeed, if we further impose that $\Delta_1=\Delta_2=\Delta_3 = d/6$, and if all interactions of a given valence $q$ are assigned one and the same coupling constant $\eta_q$, the Lagrangian is actually invariant. An example of such symmetries is
 \begin{eqnarray}
\begin{aligned}
 X \to Y\,,\, Y\to Z\,,\, Z\to \bar X\,, u \to  v\,,\, v \to  \bar w \,,\, w \to  u\,.
\end{aligned}
\end{eqnarray}

\subsubsection{Renormalisation and double-trace terms}
\label{renom_3}
Single-trace couplings associated to the interactions in $\mathcal{L}_{sT}$ are fixed at any value under the RG transformations. Indeed, one can readily convince oneself that any insertion of a chiral vertex inside its own single-trace vertex function increases the genus of the Feynman diagram by at least one unit, producing the suppression factor $1/N^2$ \cite{Caetano:2016ydc}.

In order to complete the discussion of renormalisation we should consider that $2,3$-point functions of short composite single-trace operators may get divergent quantum corrections in the loop expansion. Thus, we shall consider all the irreducible multi-trace $n$-point functions $\Gamma^{(n)}$ of the theory that have bare dimension equal to $d$ and are $U(1)^{\otimes 3}$ invariant. These quantities are responsible for the appearance of multi-trace counter-terms in the Lagrangian $\mathcal{L}_{dT}$, ubiquitous in various deformations of SYM theories~\cite{Dymarsky:2005uh,Pomoni:2009joh,Tseytlin:1999ii},  and therefore of running couplings that break conformality.

For generic angles in the loom, i.e. generic scaling dimensions of the fields, the aforementioned requirements are met only by vertices obtained via splitting of single-trace interactions \eqref{sextic}-\eqref{quartic} into multi-traces. Cubic vertices do not matter here, because any such splitting of a cubic vertex would involve at least one trace of a single field, which is zero due to $SU(N)$ symmetry.

Starting from $\Gamma^{(4)}$ we shall consider the first class of double-trace vertices given by those double-trace splitting of the vertices \eqref{quartic} that contain inside each trace both a field -- say $\Phi=\{X,Y,Z\}$ -- and its dual -- say $\phi=\{u,v,w\}$.
That is, we shall consider the vertices 
\begin{equation}
\label{dT_classI_counter}
 \text{Tr} \left[\Phi \phi \right]  \text{Tr} \left[\bar{\Phi} \bar{\phi}\, \right]   \,,\, \text{Tr} \left[\Phi \bar{\phi} \right]  \text{Tr} \left[\bar{\Phi} {\phi}\, \right] \,,\,  \text{Tr} \left[\Phi \phi \right]  \text{Tr} \left[\bar{\Phi}' \bar{\phi}'\, \right]\,,\,  \text{Tr} \left[\bar \Phi \bar \phi \right]  \text{Tr} \left[{\Phi}' {\phi}'\, \right] \,.
\end{equation}
For example, let the fields be $\Phi=X,\,\phi=u$ and $\Phi'=Z,\,\phi'=w$ and let us denote the relevant single-trace couplings as follows
\begin{equation}
\eta_{4,1}^2 \text{Tr}\left[X u \bX \bar u \right]\,,\, \eta_{4,3}^2 \text{Tr}\left[w Z \bar w \bZ \right]\,, \eta_{4,13}^2 \text{Tr}\left[X u \bar w \bar Z \right]\,,\, \bar{\eta}_{4,13}^2 \text{Tr}\left[\bar X \bar u w Z \right]\,.
\end{equation}
and similarly for double-trace counter-terms to be subtracted to the Lagrangian:
\begin{equation}
\alpha_{4,1}^2 \text{Tr}\left[X u \right] \text{Tr} \left[\bX \bar u \right]\,,\, \alpha_{4,3}^2 \text{Tr}\left[w Z\right] \text{Tr}\left[ \bar w \bZ \right]\,, \alpha_{4,13}^2 \text{Tr}\left[X u\right]   \text{Tr}\left[ \bar w \bar Z \right]\,,\, \bar{\alpha}_{4,13}^2 \text{Tr}\left[\bar X \bar u \right]  \text{Tr} \left[ w Z \right]\,.
\label{dt_quartic}
\end{equation}
The vertex $\text{Tr}\left[X u\right] \text{Tr}\left[\bX \bar u \right]$ receives divergent contributions from the series of bubble-diagrams in the loop expansion:
\begin{center}
\includegraphics[scale=0.85]{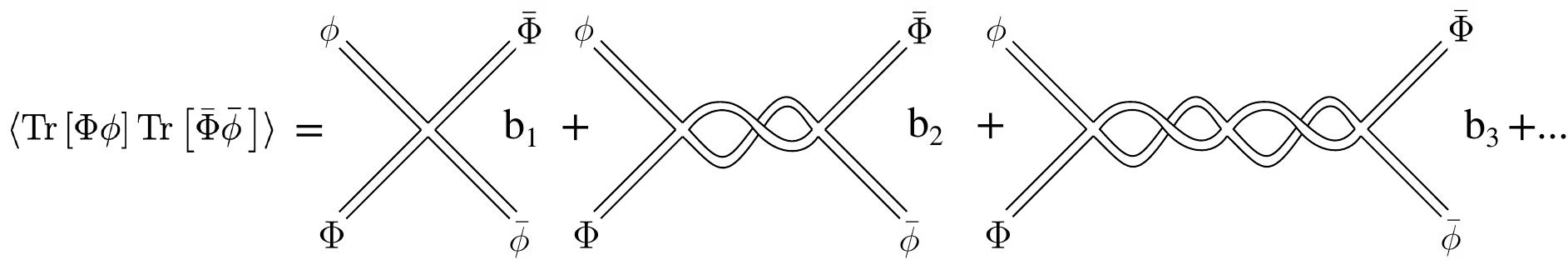}
\end{center}
The coefficients $\mathbf b_k$ are functions of the coupling constants. Omitting the numerical factors for each diagram, we can write them as
\begin{align*}
\begin{aligned}
&\mathbf b_1 = \eta^2_{4,1}-\alpha_{4,1}^2\,,\, \mathbf b_2 = (\eta^2_{4,1}- \alpha_{4,1}^2)^2+|\eta^2_{4,13}- \alpha_{4,13}^2|^2 \, , \\
&\mathbf b_3 = (\eta^2_{4,1}- \alpha_{4,1}^2)^3 + 2 |\eta^2_{4,13}- \alpha_{4,13}^2|^2( \eta^2_{4,1}-  \alpha_{4,1}^2) +\!(\eta^2_{4,3}-\alpha_{4,3}^2) |\eta^2_{4,13}- \alpha_{4,13}^2|^2  \,.
\end{aligned}
\end{align*}
In a similar fashion, the vertex $\text{Tr}\left[X u\right] \text{Tr}\left[Z w \right]$ receives divergent quantum corrections from the series of bubble-diagrams:
\begin{center}
\includegraphics[scale=0.85]{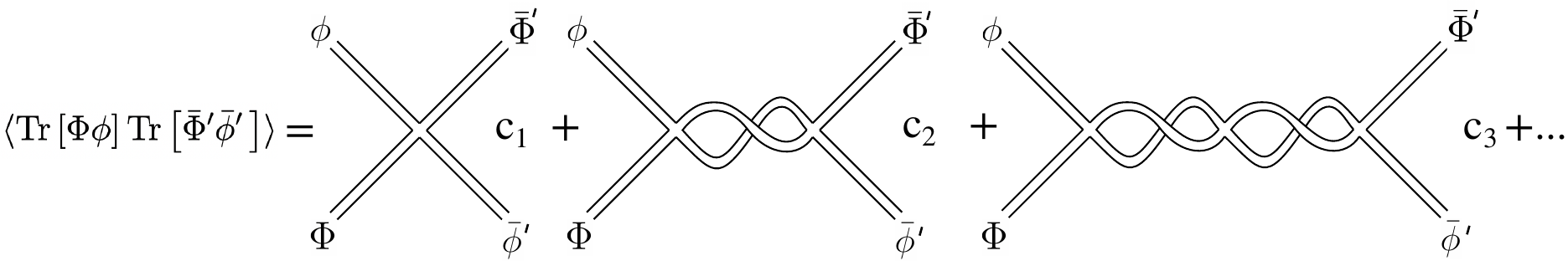}
\end{center}
where the coefficients $\mathbf c_k$ are also simple polynomials of order $k$ in the couplings.

The coefficients $\mathbf{b}_k$ and $\mathbf{c}_k$ make it transparent that the operators such as $\text{Tr}\left[Xu\right]$ and $\text{Tr}\left[Z w\right]$ -- having the same $U(1)$ charge -- can  mix at the quantum level. Moreover, the critical value for the RG flow of double-trace couplings corresponds to protected operators $\text{Tr}\left[\Phi \phi\right]$:
\begin{equation}
\label{critical_eta}
\alpha_{4,k} = \pm \,\eta_{4,k}\,,\, \alpha_{4,hk} = \pm  \,\eta_{4,hk}\,.
\end{equation}
There is actually a second class of double-trace splittings of \eqref{quartic} vertices, namely
\begin{equation}
\label{dt_finite}
 \text{Tr}\left[\Phi \phi' \right] \text{Tr}\left[\bar \Phi' \bar  \phi  \right]\,,\,  \text{Tr}\left[\Phi \bar \phi' \right] \text{Tr}\left[\bar \Phi'   \phi  \right]\,.
 \end{equation}
The quantum corrections to these vertices are finite, since the sum of the powers of propagators in each bubble is in general different from $(d/2 \mod \,\mathbb{N})$.~\footnote{Nevertheless, for special values of the angles in the loom also these bubbles become divergent and they generate double-trace couplings in the action.} Therefore, no further quartic double-trace counter-terms are needed.
Let us then consider the  double-trace vertices $\Gamma^{(5)}$ that receive divergent quantum corrections, namely
\begin{equation}
\label{dT_quintic}
\text{Tr}\left[ ABC \right] \text{Tr}\left[\Phi \phi \right]\,,\, \text{Tr}\left[ABC \right] \text{Tr}\left[\Phi \bar \phi \, \right],
\end{equation}
where $A,B,C$ are three fields that one can read out of \eqref{quintic}. The infinities in the loop expansion are removed by quartic double-traces \eqref{dt_quartic} that renormalise the series of bubbles at any loop. It is left to study double-traces $\Gamma^{(6)}$, namely
\begin{equation}
\alpha_{6,1} ^2\text{Tr}\left[XYZ \right] \text{Tr}\left[\bar X\bar Y\bar Z \right]\,,\, \alpha_{6,2} ^2\text{Tr}\left[YZ\bar X \right] \text{Tr}\left[\bar Y\bar Z X \right]\,,\,  \alpha_{6,3} ^2\text{Tr}\left[Z\bar X \bar Y \right] \text{Tr}\left[\bar Z X Y \right]\,.
\end{equation}
The same observation continue to hold in this case as well:  the renormalisation of double-trace vertices is given by a series of quartic bubbles of the type field/(dual~field) $\Phi\phi$. It is indeed not possible to create a new series of planar corrections intertwining three propagators, as it is clear from the following picture:
\begin{center}
\includegraphics[scale=0.8]{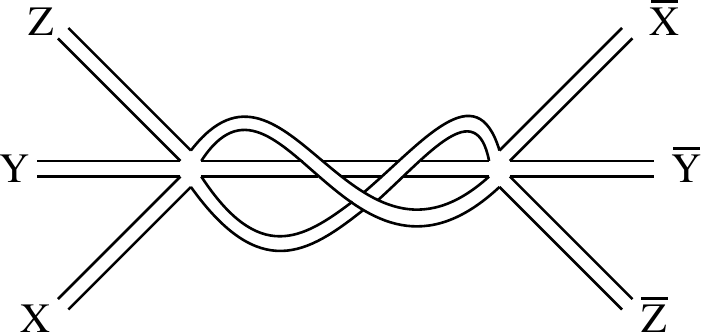}
\end{center}
Each $\Gamma^{(6)}$ stays finite under the renormalisation of quartic double-traces, with  except  the two-loop bubble that contains quintic vertices only:
\begin{center}
\includegraphics[scale=0.8]{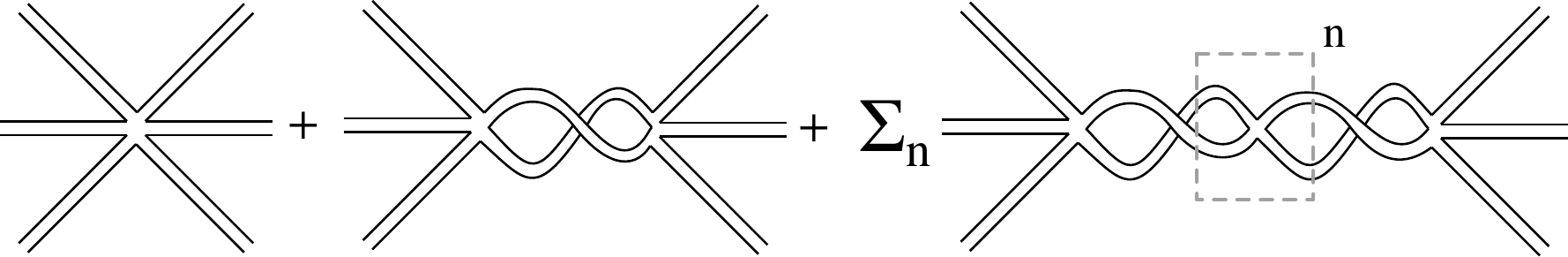}
\end{center}
Hence, at the point \eqref{critical_eta} the couplings $\alpha_{6,k}^2$ receive counter-terms at two loops only, with the renormalization factor
\begin{equation}
 (Z_{\alpha_{6,k}^2} -1) \propto \frac{1}{\epsilon} \frac{|\eta_{5,k}^2-\alpha_{5,k}^2|^2}{\alpha_{6,k}^2}\,.
\end{equation}
Let us consider for instance the vertex $\text{Tr}\left[XYZ \right] \text{Tr}\left[\bar X\bar Y\bar Z \right]$; the Callan-Symanzik $\beta$-function at one loop reads
\begin{equation}
\beta_{6,1} \propto \frac{|\eta_{5,1}^2 -\alpha_{5,1}^2|^2 }{\alpha_{6,1}^2}\,,
\end{equation}
and in order to restore scale-invariance at the quantum level we add, together with  \eqref{critical_eta},  additional criticality conditions
\begin{equation}
\label{critical_eta_5}
\alpha_{5,k} = \pm  \eta_{5,k}\,.
\end{equation}
At the critical point, the sextic double-trace couplings are zero. The operator $\text{Tr}\left[XYZ\right]$ is not protected but it receives anomalous dimension at the order $\eta_{6,1}^2$ only, due to a $1/\epsilon$-divergent bubble graph. This fact makes $\langle \text{Tr}\left[XYZ\right]  \text{Tr}\left[\bar X\bar Y\bar Z\right] \rangle$ the only 
non-protected in the mixing among operators 
\begin{equation}
\{\text{Tr}\left[XYZ\right], \text{Tr}\left[Xu\right],\text{Tr}\left[wZ\right]\}\,,
\end{equation}
defined by $U(1)^{\otimes 3}$ charge $(1,1,1)$ and bare dimension-$d/2$.
\noindent

\begin{figure}[h]
\begin{center}
\includegraphics[scale=0.8]{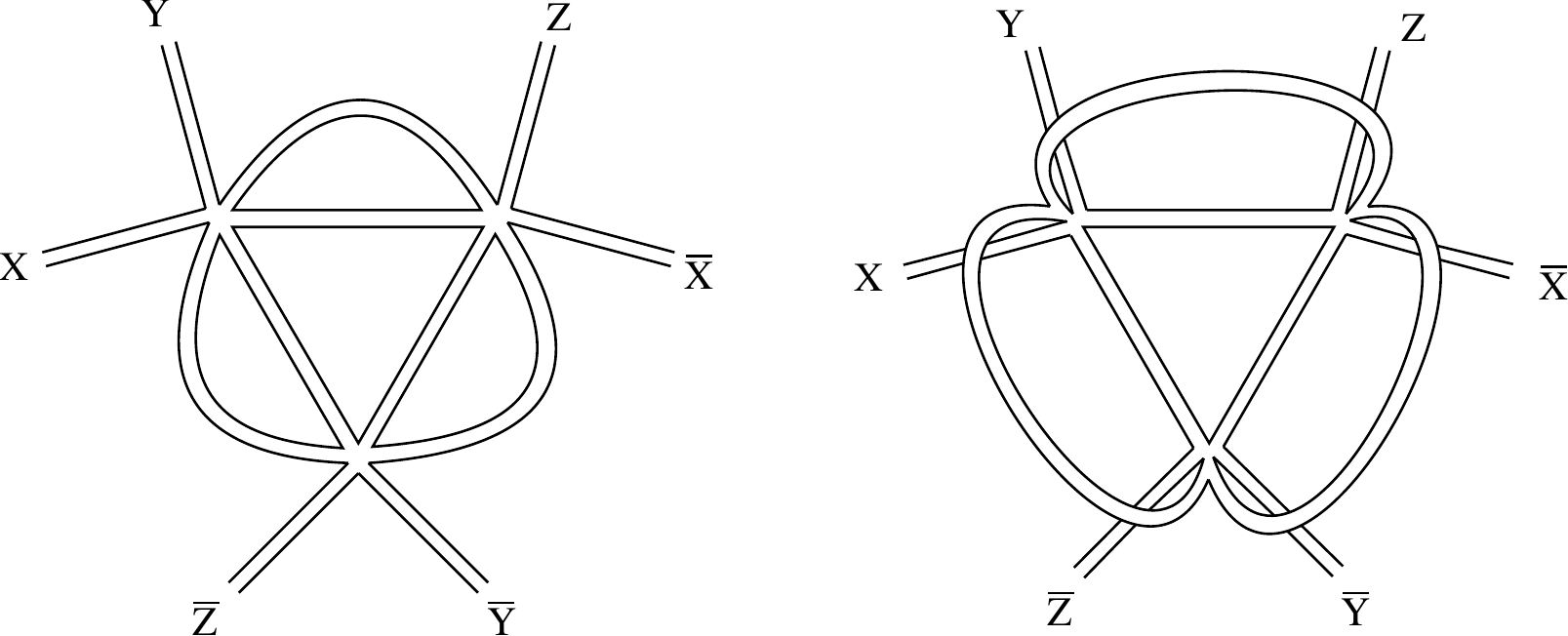}
\end{center}
\caption{The irreducible quantum corrections to the three-point functions $\text{Tr}\left[(XY) (Z \bar X) (\bar Y \bar Z) \right]$ and $\text{Tr}\left[XY\right]\text{Tr}\left[Z \bar X \right]\text{Tr}\left[\bar Y \bar Z\right]$. Both contributions are realised by vertices with the wrong chirality and cannot appear at leading order in $N$ in the loom FCFTs.}
\label{triple_traces}
\end{figure}

The multi-trace splitting of \eqref{sextic} can generate also triple-trace vertices, for instance $\text{Tr}\left[X Y\right]\text{Tr}\left[Z \bar X\right]\text{Tr}\left[\bar Y \bar Z \right]$, which may need to be renormalised. We shall therefore look at $3$-point functions of length-$2$ single-traces. As we exemplify in figure \ref{triple_traces} the $2\text{PI}$ corrections at leading order in $N$ violate the chirality of the sextic vertices involved, and we can conclude that no counter-term is needed.

Finally, let us point out that the discussion of renormalisation becomes richer once we set the angles in the Baxter lattice to some special points. For instance, whenever $\Delta_{\Phi} =\Delta_{\Phi'}$ the bubble corrections to vertices of type \eqref{dt_finite} become divergent and need to be renormalised similarly to vertices \eqref{dT_classI_counter}. More interestingly, one may specify the lattice angles so to generate wrapping divergencies in double-trace terms like $\text{Tr}\left[X^2 \right]\text{Tr}\left[\bar X^2\right]$. In $d$-dimensions these counter-terms are needed iff $[X]=d/4$, which is the condition $\theta_{12}=\pi/2$ on the angles of the Baxter lattice. 
The occurrence of this class of double-traces is very constrained: for example when $2\theta_{12}+\theta_{23}=\pi$ the theory features the double-trace counter-term $\text{Tr}\left[X^2Y\right]\text{Tr}\left[\bar X^2 \bar Y\right]$, but a theory that features both of above counter-terms would require $\theta_{23}=0$, that is a degeneration of the Baxter lattice. The wheel-graphs contributing to these two-point correlators are depicted on figure~\ref{alphaXX}.
\begin{figure}
\includegraphics[scale=1.1]{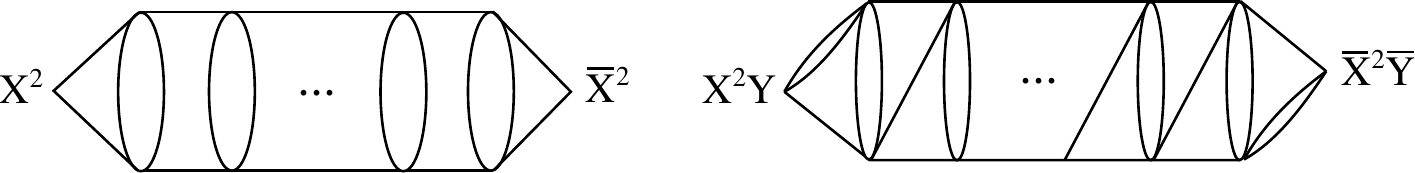}
\caption{The generic Feynman integral that corrects the two-point function of $\text{Tr}\left[X^2\right]$ by wrappings around the cylinder (left) and generates double-trace counter-terms at $\theta_{12}=\pi/2$. The analogue quantities for the two-point function of $\text{Tr}\left[X^2Y\right]$ has a similar iterative structure of wrappings, but with valence-5 vertices (right).}
\label{wrappings_XY}
\end{figure}

The critical value for the coupling $\alpha_{XX}^2$ that renormalises the operator $\text{Tr}[X^2]$ when $\theta_{12}=\pi/2$ can be computed in perturbation theory. Following the appendix B of \cite{Gromov:2018hut}, the Callan-Symanzik $\beta$-function shall take the form of a quadratic polynomial at every order in perturbation theory \cite{Pomoni:2009joh}.  Its coefficients depend on the coupling $\eta_4^2$ associated to $\text{Tr}[X u \bar X \bar u]$:
\begin{align}
\label{alphaXX}
\begin{aligned}
\beta \left(\alpha_{XX}^2 \right)=  f_2(\eta_4^2) \alpha_{XX}^4 + f_1(\eta_4^2)\alpha_{XX}^2 +  f_0(\eta_4^2)\,.
\end{aligned}
\end{align}
The coupling $\alpha_{XX}^2$ is well defined as a function of $\eta_4^2$ at the two, complex conjugated, critical values -- the zeros of the \(\beta\)-function. Similar arguments hold about the existence of critical values for the coupling $\alpha_{XXY}^2$ that renormalises the operator $\text{Tr}[X^2Y]$ in the theory with $2\theta_{12}+\theta_{23}=\pi$.

\subsection{Loom FCFT with 4 slopes}
\label{sec:4loom}

In order to formulate the general \(M\)-Loom FCFT it is essential to analyse the case of $M=4$ slopes in detail. Notably, for $M>3$ the structure of the Feynman diagrams cannot be put into that of a lattice of highest(lowest)-valence vertices, as it is the case of the triangular(honeycomb) lattice for $M=3$. In fact, it turns out that a replacement rule similar to \eqref{replacement} is enough to generate all interactions, i.e. the vertices appearing in the Baxter loom can still be obtained by star-triangle transformations starting from the $2M=8$-valence vertex.

The field content of $4$-loom FCFT consists of $M(M-1)=12$ complex scalars. We find it convenient to denote four of them as
\begin{equation}
X_1\,, X_2\,, X_3\,, X_4\,,
\end{equation}
with dimensions $\Delta_1,\Delta_2,\Delta_3,\Delta_4$, and then the fields with ``dual" scaling dimensions \(\Delta'_i=d/2-\Delta_i\) as
\begin{equation}
u_1\,, u_2\,, u_3\,, u_4\,.
\end{equation}
We denote the last four fields as
\begin{equation}
Y_1,Y_2\,,\qquad \text{and their duals}\quad v_1,v_2\,,
\end{equation}
with scaling dimensions $\Delta_5,\Delta_6$ and  \(d/2-\Delta_5 ,\,d/2-\Delta_6\), respectively. The dimensions of fields are subject to the constraints imposed by the scale invariance of the interaction vertices -- the sum of dimensions of propagators around any vertex is always equal to \(d\) -- which is built-in via the geometry of closed polygons in the dual, loom graph construction (see section~\ref{sec:graphs}).

\begin{figure}[t]
\begin{center}
\includegraphics[scale=1.2]{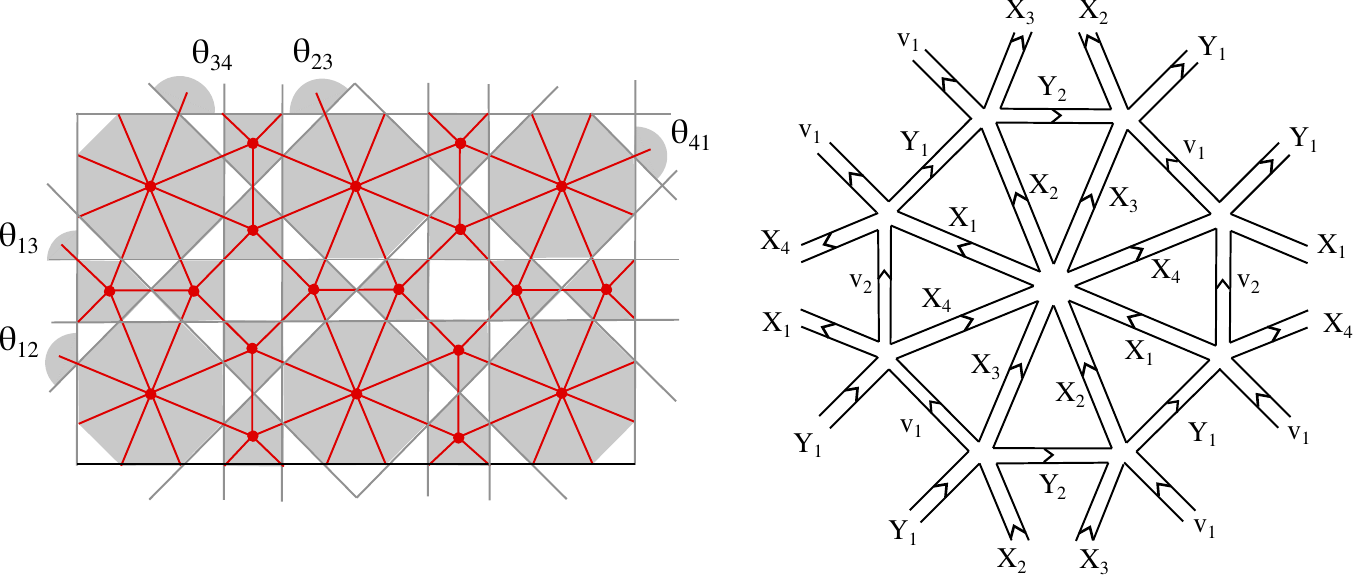}
\caption{Left: a portion of Baxter lattice made of lines with slopes numbered from $1$ to $4$ and the associated Feynman diagram. We highlight (some of the) different angles appearing between slopes and corresponding to different propagators/fields in the theory. Right: the highest valence vertex \eqref{8_vert} and the surrounding vertices in detail, in double-line notation.}
\label{4Loom_V}
\end{center}
\end{figure}

The fields interact via the total of $131$ single-trace chiral vertices. The highest valence facet one can build in the loom is an octagon with four couples of parallel edges, and the corresponding valence-$8$ vertex represented in figure \ref{4Loom_V} (right) reads
\begin{align}
\label{8_vert}
\text{Tr}\left[X_1 X_2 X_3 X_4 \bar X_1 \bar X_2 \bar X_3 \bar X_4 \right]\,,
\end{align}
compatible with the constraint $\Delta_1+\Delta_2+\Delta_3+\Delta_4 = \tfrac{d}{2}$.
\noindent
All vertices of valence $7$ can be obtained by one star-triangle move starting from \eqref{8_vert}, replacing two fields $X_i$ with one field $Y_i$ or $v_i$ according to the rules
\begin{equation}
\label{rep_4_STR}
X_1 X_2 \to v_1\,,\, X_2 X_3 \to v_2 \,,\, X_3 X_4 \to Y_1\,,\, X_4 \bar X_1 \to Y_2\,,\, 
\end{equation}
and thus reducing an octagonal loom face to a heptagonal one in $8$ different ways:
\begin{align}
\begin{aligned}
&\text{Tr}\left[v_1 X_3 X_4 \bar X_1 \bar X_2 \bar X_3 \bar X_4 \right]\,,\, \text{Tr}\left[X_1 v_2 X_4 \bar X_1 \bar X_2 \bar X_3 \bar X_4 \right] \,,\, \text{Tr}\left[X_1 X_2 Y_1 \bar X_1 \bar X_2 \bar X_3 \bar X_4 \right]\\
& \text{Tr}\left[X_1 X_2 X_3 Y_2 \bar X_2 \bar X_3 \bar X_4 \right]\,,\, \text{Tr}\left[X_1 X_2 X_3 X_4 \bar v_1 \bar X_3 \bar X_4 \right] \,,\, \text{Tr}\left[X_1 X_2 X_3 X_4 \bar X_1 \bar v_2 \bar X_4 \right]\\
& \text{Tr}\left[X_1 X_2 X_3 X_4  \bar X_1  \bar X_2 \bar Y_1 \right] \,,\,  \text{Tr}\left[X_2 X_3 X_4 \bar X_1 \bar X_2 \bar X_3 \bar Y_2 \right]\,. \\
\end{aligned}
\end{align}
As we said, there is no planar Feynman diagram with only valence-$8$ vertices. The closest relative to the Kagom\'{e} lattice of $M=3$ is the tiling of octagons and pentagons showed in the figure \ref{4Loom_V}. There are four  valence-5 vertices appearing in the diagram of figure \ref{4Loom_V}, namely
\begin{align}
\begin{aligned}
&\text{Tr}\left[X_1  v_2 X_4 \bar v_1 \bar Y_1 \right]\,,\, \text{Tr}\left[\bar X_1 \bar v_2 \bar X_4 v_1 Y_1 \right]\,,\,   \text{Tr}\left[X_2 Y_1 \bar v_1 \bar X_3 \bar Y_2 \right] \,,\, \text{Tr}\left[\bar X_2 \bar Y_1  v_1 X_3 Y_2 \right] \,.
\end{aligned}
\end{align}
The latter are not all the vertices of valence $5$ in the Loom FCFT\(^{(4)}\): there are in total  $48$ vertices of valence five. This theory also features $28$ vertices of valence six, $38$ vertices of valence four and $8$ cubic vertices, namely:
\begin{align}
\begin{aligned}
&\text{Tr}\left[ u_1 \bar u_2 \bar v_1\right],
\text{Tr}\left[ u_1 \bar u_4 \bar Y_2\right],
\text{Tr}\left[ \bar u_3 \bar u_4 Y_1\right],
\text{Tr}\left[ u_2 \bar u_3 \bar v_2 \right], \\
&\text{Tr}\left[ u_2 \bar u_1 v_1 \right],
\text{Tr}\left[ u_4 \bar u_1 Y_2 \right],
\text{Tr}\left[ u_3 u_4 \bar Y_1 \right],
\text{Tr}\left[ u_3 \bar u_2 v_2\right]\,.
\end{aligned}
\end{align}
The complete list of the vertices of $4$-Loom FCFT is given in the appendix \ref{app:vertices}.

Leaving only a certain small number of constants non-zero in 4-Loom FCFT, we can end up with a few interesting particular FCFTs. In particular, retaining only the quartic couplings $\Tr\left[v_1 X_3 Y_2 \bar{u}_1 \right]$ and $\Tr\left[u_1 \bar{v}_1 \bar{X}_3 \bar{Y}_2 \right]$
we arrive at the  FCFT called ``Checkerboard", which contains the known FCFTs as particular cases (bi-scalar, ABJM FCFTs considered above, or BFKL-type FCFT) and generalises them to the presence of spectral parameter in the Feynman graphs~\cite{Checkerboard:2022}.

\subsubsection{$U(1)$ symmetry} 
\label{sec:U(1)M4}

Let us argue that the analogue of the $U(1)^{\otimes M}$ symmetry featured by the $M=2$ or $M=3$ loom FCFTs exists for $M=4$. We are looking for a symmetry of the type
$$\underbrace{U(1)\otimes U(1) \otimes \cdots \otimes U(1)}_{\text{\normalfont $n$-fold}},$$
such that any field, say $\Phi$, transforms acquiring a phase that depends on a set of $k$ charges $q_{\Phi,k} \in \mathbb{R}$ -- one for each copy of $U(1)$
\begin{align}
\Phi \to e^{i \sum_{k=1}^n q_{\Phi,k} \theta_k} \times \Phi\,,\,\,\,\,\,
\bar \Phi \to e^{-i \sum_{k=1}^n q_{\Phi,k} \theta_k} \times \bar \Phi\,.
\end{align}
In order to detect the symmetry, let's call $x(\Phi) = \sum^n_{k=1} q_{\Phi,k} \theta_k$ the angle associated by a given transformation for the field $\Phi$. The values of $x(\Phi)$ -- for every transformation $\{\theta_1,\dots, \theta_n\}$-- shall satisfy the same set of linear equations ensuring that each interaction vertex is invariant. Namely, given a vertex $\text{Tr}\left[\Phi_{a_1} \dots \Phi_{a_r} \right]$, the associated constraint on charges is
\begin{equation}
x(\Phi_{a_1})+\dots + x(\Phi_{a_r}) =0\, \Longleftrightarrow \,   \sum_{s=1}^r q_{\Phi_{a_s},k} = 0 \,,\,\,\, \forall k=1,\dots ,n\,.
\end{equation}
Since in the loom FCFTs the interactions appear always in pairs of the type
\beq
\text{Tr}\left[A B C\cdots \right]\,,\,\, \text{Tr}\left[\bar A\bar  B\bar  C\cdots \right]\,,
\eeq
and the charge of a field acquires a $``-"$ sign under hermitian conjugation, the system features pairs of identical equations, hence half of them must be dropped. Due to the huge number of vertices in the theory it is convenient to use $Mathematica$ to solve the system. It follows from the solution that all $x(\Phi)$ can be expressed via a linear combination of four of them $x(X_k)$. It appears that the theory is invariant under four copies of $U(1)$, and labelling the corresponding charges as $(q_1,q_2,q_3,q_4)$ we obtain
 \label{tab_charge_4}
  \begin{center}
    \begin{tabular}{c|c|c|c|c|} 
       &$q_1$ & $q_2$ & $q_3$& $q_4$ \\
      \hline
      $X_1$ & 1 & 0 & 0 &0\\
      $X_2$ & 0 & 1 &  0 &0\\
      $X_3$ & 0 & 0&1& 0\\
       $X_4$ & 0 & 0& 0&1
             \end{tabular}\;\;\; ; \;\;\;
         \begin{tabular}{c|c|c|c|c|} 
       &$q_1$ & $q_2$ & $q_3$& $q_4$ \\
      \hline
      $u_1$ & 0 & 1 & 1 &1\\
      $u_2$ & -1 & 0 & 1 &1\\
      $u_3$ & -1 & -1 &0 & 1\\
       $u_4$ & 1 & 1& 1& 0
             \end{tabular} 
            \;\;\; ; \;\;\;
         \begin{tabular}{c|c|c|c|c|} 
       &$q_1$ & $q_2$ & $q_3$& $q_4$ \\
      \hline
      $Y_1$ & 0 & 0 & 1 &1\\
      $Y_2$ & -1 & 0 & 0 &1\\
      $v_1$ & 1 & 1 &0 & 0\\
       $v_2$ &  0& 1& 1& 0
             \end{tabular}\,\,\,.
            \end{center}

The classification of fields and vertices and the analysis of $U(1)$ charges make it transparent that the interactions generated by the Loom can be read out from the highest-valence vertex \eqref{8_vert} via a rule that substitutes strings of $X_i$'s with one field
\begin{align}
\begin{aligned}
\label{replacement_4}
&X_1 X_2 \to v_1\,,\, X_2 X_3 \to v_2 \,,\, X_3 X_4 \to Y_1\,,\, X_4 \bar X_1 \to Y_2\,,\\
&X_2 X_3 X_4 \to u_1\,,\, X_3 X_4 \bar X_1 \to u_2 \,,\, X_4 \bar X_1 \bar X_2 \to u_3\,,\, \bar X_1 \bar X_2 \bar X_3 \to u_4\,.
\end{aligned}
\end{align}
Each of these substitutions preserves the scaling dimension and the $U(1)$ charges. The first line in \eqref{replacement_4} is just the star-triangle replacement of two fields with one, noticed in \eqref{rep_4_STR}, but also the second line can be broken up into star-triangles, making manifest the origin of the replacement as a feature of the Loom. Indeed, applying  the star-triangle transformations in the first line of \eqref{replacement_4} to the states in the second line, we can rewrite the latter line as follows
\begin{align}
\begin{aligned}
\label{replacement_4_STR_II}
&X_2 X_3 X_4 \to v_2 X_4 \to u_1\,,\, X_3 X_4 \bar X_1 \to Y_1 \bar X_1 \to u_2 \,,\, \\ & X_4 \bar X_1 \bar X_2 \to Y_2 \bar X_2 \to u_3\,,\, \bar X_1 \bar X_2 \bar X_3 \to \bar v_1 \bar X_3 \to u_4\,.
\end{aligned}
\end{align}
            
After the discussion of the Loom FCFTs for $M=3,4$ we are ready to proceed with the general construction for any number $M$ of slopes in the Baxter lattice, including the renormalisation and double-trace terms. 


\subsection{Loom FCFT with \(M\) slopes}

\begin{figure}[h]
\begin{center}
\includegraphics[scale=0.75]{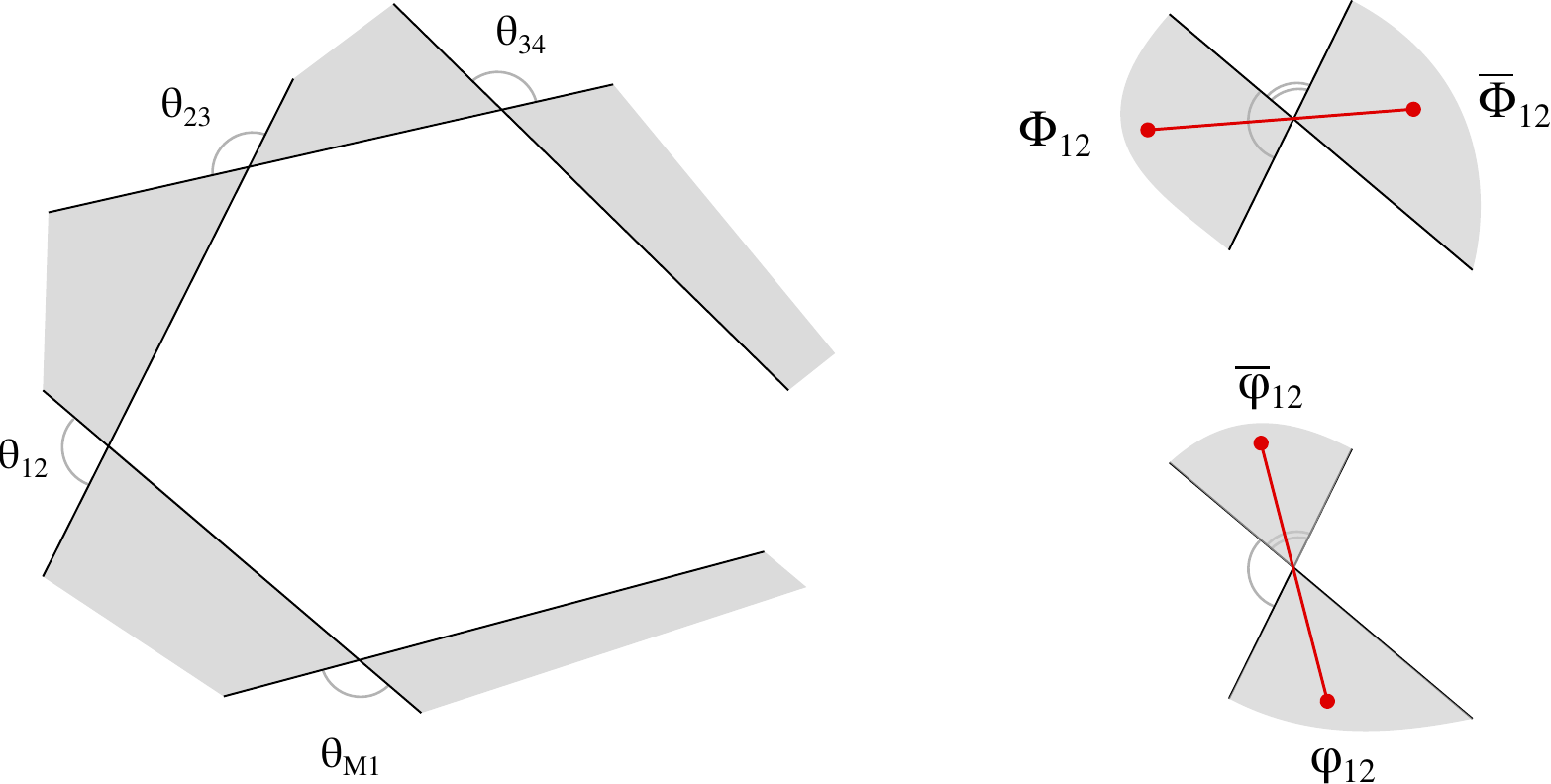}
\caption{The slopes $\{M,1,2,3,4\}$ of a loom with $M>4$, and the associated angles $\theta_{ij}$ they form on the Baxter lattice (its fragment is on the left). On the right, the Feynman diagram propagators that cross a certain Baxter lattice vertex in orthogonal directions: for fields $\Phi$ and for dual fields $\phi$.}
\label{fig:Mloom}
\end{center}
\end{figure}

In the following we present the general scheme for the construction of  Lagrangian of the Loom FCFT with any number of slopes $M$ in its the Baxter lattice. We will label the $M$ slopes as $k=1,\dots, M$. The crossing of two lines along the $h$-th and $k$-th slope of the Baxter lattice forms two complementary angles that we denote $\theta_{hk}$ and $\theta_{hk}'={\pi}- \theta_{hk}$. A lattice with $M$ slopes can generate faces with any number $n$ of edges inside the range $3\leq n \leq 2M$ by means of the star-triangle moves of the Loom, starting from simpler configurations. 
Therefore, the Feynman diagrams associated with the general Baxter lattice can feature  $M(M-1)/2$ different propagators that pass through the angles $\theta_{hk}$ - associated to complex scalar fields $\Phi_{hk}$ -- and $M(M-1)/2$ propagators that pass through the angles $\theta_{hk}'$, associated to the ``dual" fields $\phi_{hk}$, as shown in Fig.\ref{fig:Mloom}.

In order to fix the notation we introduce the ordering of slopes in the following way: we call $0<\theta_{hk}<\pi$ an angle that is comprised between the slope $h$ and the slope $k<h$ in clockwise order and $\theta_{kh}=-\theta_{hk} \in (-\pi,0)$. Unbarred and barred fields $\Phi_{hk}$ cross the same  vertex of the Baxter lattice from the side of $\theta_{hk}>0$ and $\theta_{kh}<0$ respectively. The analogous picture holds for $\phi_{hk}$ unbarred/barred fields with respect to the angle $\theta_{hk}'$.

\noindent
The fields $\Phi_{hk}$ have the scaling dimensions 
\begin{equation}
\Delta_{hk} = \frac{\theta_{hk}'}{2 \pi} d\,,
\end{equation}
while the fields $\phi_{hk}$ have the scaling dimension
\begin{equation}
\Delta'_{hk} =\frac{ \theta_{hk}}{2 \pi} d=  \frac{d}{2}- \Delta_{hk}\,.
\end{equation}
The fields $\phi$ are ``dual'' to the fields $\Phi$ in the sense that the effect of a star-triangle duality is to replace the propagators of each \(\Phi\)-field with the propagators of the corresponding \(\phi\)-field.

The Lagrangian of the loom field theory with $M$ directions has therefore a kinetic term given by the sum of a $M(M-1)$ free fields with bare dimensions $\Delta_{hk}$ and $\Delta_{hk}'$
\begin{equation}
\mathcal{L}_0 = \frac{1}{N}\sum_{h>k}^M \text{Tr}\left[ \bar{\Phi}_{hk}\left(-\partial_{\mu}\partial^{\mu}\right)^{\Delta_{hk}'} \Phi_{hk} + \bar{\phi}_{hk}\left(-\partial_{\mu}\partial^{\mu}\right)^{\Delta_{hk}} \phi_{hk} \right]\,,
\end{equation}
and it is in general non-local as it features the integral operators 
\begin{equation}
\left(-\partial_{\mu}\partial^{\mu}\right)^{a} f(x)= \pi^{\frac{d}{2}} \frac{\Gamma\left(\frac{d}{2}-a \right)}{\Gamma \left(a\right)} \int \frac{d^d y}{(x-y)^{2 \left(\frac{d}{2}-a\right)}} f(y)\,.
\end{equation}
The latter reduces to a differential one -- therefore local -- only for integer values of $``a"$. However, we stress that the locality property of the Lagrangian in a CFT is not so fundamental as for the massive theories.  

The vertices of the theory can be read out of the general Baxter lattice, analysing the possible faces one can form with lines along $M$ slopes. There is a unique vertex of maximal valence $2M$ that corresponds to a $2M$-gon on the Baxter lattice, and reads
\begin{equation}
 \frac{1}{N}\text{Tr}\left[\Phi_{12}\Phi_{23}\dots \Phi_{M 1} \bar{\Phi}_{12}\dots \bar{\Phi}_{M 1} \right](x)\,,
 \end{equation}
where $\bar{\Phi}_{hk}$ is the hermitian conjugate of $\Phi_{hk}$.~\footnote{Since the theory is non-unitary, the barred and unbarred fields can be often considered as independent variables in the functional integrals defining physical quantities} 
The interactions in the Loom FCFT can be packed into a single-trace Lagrangian
\begin{equation}
\mathcal{L}_{int} = \frac{1}{N}\sum_{n=3}^{2M}\sum_{k=1}^{\#(M,n)} \eta_{n,k}^2\, \text{Tr}\left[\dots  \Phi \dots \phi\dots  \right]\,,
\end{equation}
where $\#(M,n)$ is the number of vertices of valence $n$ for a given $M$ number of slopes in the Baxter lattice.
We stress that in the general Loom FCFT each single trace vertex can enter  with an independent coupling which we denoted as $\eta_{n,k}^2$. However, we will impose the additional requirement that each pair of vertices of the type
\begin{equation} 
\label{conjugate_Vertices}
 \eta^2 \, \text{Tr}\left[A BC\dots F G \right] + \bar \eta^2 \, \text{Tr}\left[\bar A \bar B \bar C\dots \bar F \bar G \right]\,,
\end{equation}
enters with complex conjugate couplings $\bar \eta =(\eta)^*$. This will impose natural restictions to the spectrum of anomalous dimensions, as it will be discussed in section~\ref{sec:tTsym}.

\subsubsection{Counting of vertices in FCFT with \(M\) slopes}
The counting and classification of vertices becomes quickly cumbersome as $M$ increases. In order to count the vertices of a given valence $n$ one should count the number of the ways by which one can construct an $n$-gon having the edges along a subset of slopes of the loom. This problem can be rephrased in terms of ordered partitions. Given a slope $h$ we can assign it two elements $h$ and $\bar h$ that are its two orientations. A polygon with $n$ edges corresponds to an ordered subset of $n$ elements from the string
\begin{equation}
\label{seq_M}
1\,2\,3\,\dots M\,\bar 1\,\bar 2\, \dots \, \bar M\,,
\end{equation}
such that its first element $i$ and its last element is $\bar j$ with $j>i$. Given a subset $i l h k\dots \bar j$, any pair of successive elements denotes the crossing of two lines in the Baxter lattice, and thus the associated propagator/field in the vertex. For instance, one can associate $hk$, $h\bar k$ and $\bar h\bar k$ to $\Phi_{hk}$, $\phi_{hk}$ and $\bar{\Phi}_{hk}$ respectively. It is evident that  each pair of the type $h\bar k$ is constrained by $k<h$ since a given direction $h$ cannot cross itself $\bar h$ or even more tilted lines $\bar k >\bar h$. 

Each vertex of valence $n$ is a subsequence of $n$ letters extracted from \eqref{seq_M} for which we can specify the first and last ``letter" in the unbarred/barred subsets: $a \dots c\, \bar b \dots \bar d$. Let $l$ be the number of unbarred letters and $n-l$ the number of barred ones. 
Once $a$ and $c$ are fixed, there are $l-2$ intermediate letters to choose in $a+1, \dots ,c-1$, which is a set of length $a-c-1$. Similarly, there are $n-l-2$ unbarred letters to be chosen  inside $b+1,\dots,d-1$ once $b$ and $d$ are fixed. Thus once $a,c,b,d$ are fixed, for a given $l$ the number of generated vertices is
\begin{equation}
\sum_{l=1}^{c-a+1}
 \begin{pmatrix} c-a -1  \\ l -2
\end{pmatrix} \begin{pmatrix} d-b-1 \\ n-l-2
\end{pmatrix}\,.
\end{equation}
Last formula is correct only whenever $c>a$ and $b>d$. If $a=c$ one has only one choice, the unbarred string is $``a"$ and $l$ can only be zero. Similarly for $b=d$ the barred string is $``d"$ and $n-l$ shall be $0$. We can simply correct the binomials by  $\delta_{a,c}$ and $\delta_{b,d}$ shifts that take into account these latter cases too, and the formula for the number of vertices of valence \(n\) in the \(M\)-Loom FCFT reads:
 \begin{equation}
\#(n,M)= \sum_{a,b=1}^{M} \sum_{c= \text{max}(a,b)+1}^M \sum_{d=\text{max}(a,b)+1}^M \sum_{l=1}^{c-a+1} \begin{pmatrix} c-a -1  \\ l -2+\delta_{a,c}
\end{pmatrix} \begin{pmatrix} d-b-1 \\ n-l-2+\delta_{b,d}
\end{pmatrix}\,.
\end{equation}
For the first few $M$'s the number of vertices $\#(n,M)$ of valence $n \leq 12$ reads
  \begin{center}
          \begin{tabular}{l| c c c c c c c c c c c c } 
    \textbf  M & $3$ &$4$ & $5$ & $6$ & $7$ & $8$ & $9$ & $10$ & $11$ & $12$ \\
      \hline
      \textbf{3} & 2 & 9 & 6 & 1 & 0 & 0 & 0 & 0& 0 & 0 \\
    \textbf   4 & 8 & 38 & 48 & 28 & 8 & 1 & 0 & 0& 0 & 0  \\
     \textbf 5 & 20 & 110 & 202 & 200 & 120 & 45 & 10 & 1 & 0 & 0  \\
    \textbf  6 & 40 & 255 & 612 & 852 & 780 & 495 & 220 & 66 & 12 & 1  \\
    \end{tabular}
  \end{center}
 Thus, for a general Loom FCFT the number of coupling constants grows very fast (exponentially) with $M$. For instance, these numbers are $\{18, 131, 708, 3333,\dots\}$ for $M=3,4,5,6,\dots$, respectively. However, we have the freedom to restrict these couplings in an arbitrary way. Say, we can set any subset of these couplings to be equal, imposing additional symmetries on the action,  and/or  set some of them to  zero, excluding  certain interactions from the theory without spoiling the integrability of diagrams.  In this process, some  fields can disappear from interaction terms and thus can be decoupled from the theory. This is the way to construct particular reductions, useful for various physical questions,  such as bi-scalar or ABJM FCFT discussed in the previous sections.

\subsubsection{U(1) symmetry for general $M$}
The counting of vertices in the previous section can be matched against the replacement rule of the type \eqref{replacement} for $M=3$ or \eqref{replacement_4} for $M=4$, based on the star-triangle symmetry of the Loom. This rule becomes more transparent once we look at the internal $U(1)$-symmetries of the $M$-Loom FCFT. This symmetry can be spotted by solving directly a system of neutrality constraints on the  vertices of the theory, as it was done for $M=4$ in section \ref{sec:U(1)M4}, or rather derived by star-triangle moves applied to the highest-valence vertex
\begin{equation}
    \text{Tr}\left[\Phi_{12}\cdots \Phi_{M 1}\,\bar \Phi_{1 2} \cdots \bar \Phi_{M 1} \right]\,.
\end{equation}
That is, we shall consider a symmetry group
\begin{equation}
    \underbrace{U(1)\otimes U(1) \otimes \cdots \otimes U(1)}_{\text{\normalfont $M$-fold}}
\end{equation}
and associate to  $M$ fields $\Phi_{i,i+1}$ a unit-vector charge
\begin{equation}
    q(\Phi_{i,i+1}) =(0,\dots, \underbrace{1}_{\text{\normalfont $i$-th}},\dots ,0)\,.
\end{equation}
Next, replacing a couple of fields $\Phi_{i-1,i}\Phi_{i,i+1}$ in the $2M$-vertex with another Loom field $\phi$ via star-triangle, the neutrality of the newly-obtained vertex reads
\begin{equation}
    q(\phi) = q(\Phi_{i-1,i})+q(\Phi_{i,i+1})\,.
\end{equation}
The procedure can be iterated by applying other star-triangle moves to the vertex, all the way(s) until it reduces to a cubic one. By construction from the Loom, this iteration catches all the vertices in the theory and involves all the fields in it.
Nicely, it provides a way to classify fields according to the ``level", i.e. the number of star-triangle moves necessary for a field to appear. For instance, at $M=4$ the level-$0$ fields are $X_k$, then $Y_k$ and $v_k$ are level-$1$ and fields $u_k$, dual to $X_k$, are level-$2$. A field of level $n$ can be regarded as the replacement of a string of $n+1$ fields of level-$0$
\begin{equation}
    \Phi_{i,i+1} \Phi_{i+1,i+2} \cdots \Phi_{i+n,i+n+1} \to \phi \,\,(\text{level-$n$})\,.
\end{equation}
Last equation determines at once the scaling dimensions of $\phi$ and its $U(1)$-charges
\begin{equation}
    [\phi] = \sum_{j=1}^n   [\Phi_{i+j,i+j+1}]\,,\,\,\,\,\,q(\phi) = \sum_{j=1}^n   q(\Phi_{i+j,i+j+1})\,.
\end{equation}
It also provides a quick check on the number of fields in the theory compatible with the symmetries of the Loom.
Starting from $M$ complex fields at the level-$0$,  any pair $\Phi_{i-1,i} \Phi_{i,i+1}$ can be substituted by a single field, resulting in $M$ complex fields at level-$1$. The same is true all the way until the level $M-2$, making a total of $M(M-1)$ complex fields and $M(M-1)$ conjugate fields.

\subsubsection{Renormalisation and conformal symmetry of FCFTs}
\label{sec:renorm}

The fixed chirality of  interactions (each term does not have its complex conjugate) generated by the Loom construction implies that in the large-$N$, multi-color limit the single-trace vertices do not receive corrections at any order of the weak coupling expansion from graphs that contain only the single-trace interactions. Thus, in any Loom FCFT the single-trace couplings have fixed values under the renormalisation group (RG) transformations as it was first observed in the bi-scalar theory~\cite{Sieg:2016vap,Gromov:2018hut}.

Nevertheless, the single-trace interactions of the theory are not UV finite as they can generate infinities in the two-point (or higher-point) functions of single-trace operators at loop level. These divergences cannot just be multiplicatively renormalised since the single-trace interactions can generate  double-trace (or multi-trace) divergent vertex functions that require the introduction of counter-terms with generically running couplings in order to be renormalised. 

In order to analyse the renormalisation of a Loom FCFT one should understand which counter-terms can be generated, i.e. what are the possible scale-invariant and $U(1)^{\otimes M}$-invariant multi-trace vertices to be added to the interaction Lagrangian. For general angles in the Baxter lattice, this is equivalent to look for all possible multi-trace splittings of the single-trace Loom FCFT interactions that preserve the order of fields. For instance starting from $\text{Tr}\left[A B C D E F \right]$ the candidate multi-trace counter-terms are
\begin{equation}
\label{splittings}
\Big \{ \text{Tr}\left[A B C D  \right] \text{Tr}\left[E F \right],\, \text{Tr}\left[A B C \right] \text{Tr}\left[DEF \right],\, \text{Tr}\left[A B \right] \text{Tr}\left[ C D \right] \text{Tr}\left[E F \right] , \text{cyclic perm.}\Big\} 
 \end{equation}
  where all the fields $A,\dots F$ are  different -- as it is always the case in a Loom FCFT vertex.
Any reshuffling of fields inside the traces that is not equivalent to \eqref{splittings} by trace cyclicity, say for instance $ \text{Tr}\left[A B D C  \right] \text{Tr}\left[E F \right]$ or $ \text{Tr}\left[A D \right]  \text{Tr}\left[ B C \right]  \text{Tr}\left[E F \right]$, shall be a priori  excluded since they violate the chirality of the theory and thus would generate sub-leading terms in the large-$N$ expansion.

Furthermore, we can generalise two observations made at $M=3$. First, due to the chirality the only way to get divergent quantum corrections to a double-trace vertex is by means of a series of bubbles that ``intertwine" two propagators of the type ``field/dual~field": $\Phi/\phi$. On the contrary, trying to generate bubbles by intertwining more than two propagators gives contributions that are subleading at large-$N$:
\begin{center}
\includegraphics[scale=0.75]{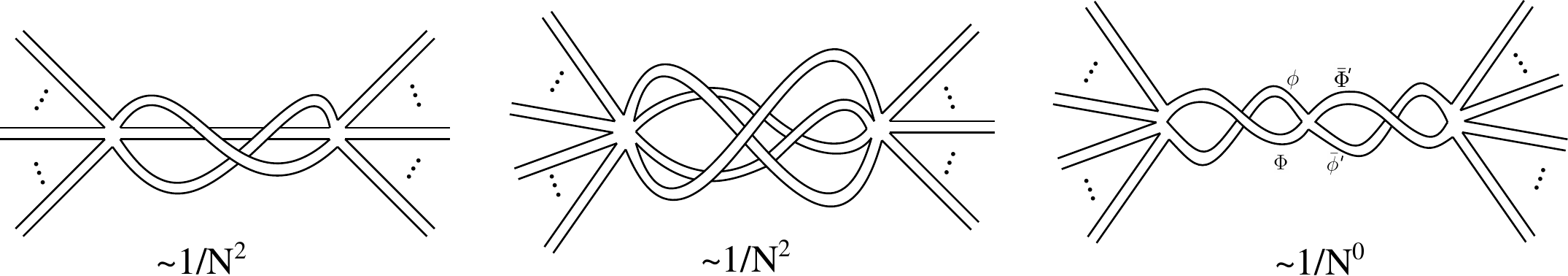}
\end{center}
Notice that the entire bubble series besides the one-bubble term is generated by the insertion of valence-4 interactions.

Secondly, there are no divergent irreducible vertex functions with more than two traces, i.e. such multi-trace vertices do not get renormalised at leading order in large-$N$ limit and should be dropped from our analysis.

On the right picture of last figure there is an example of quartic bubble inserted in a double-trace correlator with any number of external fields. The sum over such bubble insertions is divergent and actually requires the coupling renormalisation if and only if the two propagators sum up to ${d}/{2}$ (plus integers). Examples of relevant vertices in this analysis are the vertex $\text{Tr}[XY]\text{Tr}[\bar X \bar Y]$ of the bi-scalar fishnet \eqref{FCFTbiscalar}~\cite{Sieg:2016vap,Gromov:2018hut}, or  of the vertices
\begin{equation}
\text{Tr}\left[XYZ\right] \text{Tr}\left[\bar X \bar Y  \bar Z  \right]\,,\, \text{Tr}\left[XYZ\right] \text{Tr}\left[\bar X \bar u\right]\,,\, \text{Tr}\left[X u\right] \text{Tr}\left[\bar X \bar u\right]\,,\end{equation}
of $3$-Loom FCFT, as well as of the following vertices of of $4$-Loom FCFT:
 \begin{equation}
\text{Tr}\left[X_1  X_2 X_3 X_4 \right] \text{Tr}\left[\bar X_1  \bar X_2 \bar  X_3  \bar X_4 \right]\,,\, \text{Tr}\left[X_1  v_2\right]\text{Tr}\left[X_4 \bar v_1 \bar Y_1 \right]\,.
\end{equation}
A similar analysis of such terms can be reproduced for general $M$-Loom FCFT and one comes to the same conclusions as observed for $M=2,3$.
At special points in the space of loom FCFTs -- that is for special angles $\theta_{ij}$ in the Baxter lattice -- there can be other double-trace counter-terms which are induced by wrapping corrections \cite{Sieg:2016vap,Grabner:2017pgm,Gromov:2018hut}, as it was discussed in section \ref{renom_3} for $M=2,3$ with the examples of $\text{Tr}[X^2]\text{Tr}[\bar X^2]$ and $\text{Tr}[X^2Y]\text{Tr}[\bar X^2 \bar Y]$.
A classification of these second class of vertices can be done once we fix $M$ and constrain some angles at which they can appear, appropriately.
As a general conclusion, we can claim that the Loom FCFTs are UV complete and scale-invariant at the quantum level when the double-trace couplings are set to specific values -- i.e. the critical points, that achieve the conformality of field theories via the Baxter lattice construction.  However, we stress again that  our Loom FCFTs are  non-unitarity theories due to the chirality of their Lagrangian. That means that the FCFTs are genuinely logarithmic CFTs \cite{Gurarie:1993xq,Gromov:2017cja}.
 
\subsection{$d=4$ spinning FCFT}

The general construction described above  can be further generalised replacing  scalars with spinning fields. Let us focus on the definition of a spinning Baxter lattice in $d=4$ dimensions, corresponding to a loom FCFT action containing fields that transform under the product of two irreps of $SU(2)_L$ and $SU(2)_R$, e.g. left/right fermions $(\tfrac{1}{2},0)$ and $(0,\tfrac{1}{2})$. The loom picture on the Baxter lattice is modified by associating   spins to the slopes \(j=1,2,\dots,M\), such that  the $j$-th slope is defined not only by its angle but also by the $SU(2)$ representation of spin $s_j$.

Locally the Baxter lattice has the structure of crossings of straight lines, and each crossing is labelled by the two spins $(s_i,s_j)$ of the intersecting lines that define the left/right spin of the propagator that passes through this crossing. The rule associating angles to scaling dimensions of fields takes now into account also the spins:
\begin{equation}
    \Delta = \frac{\pi-\theta}{2\pi} d +s_i +s_j\,.
\end{equation}
We shall introduce fermions on the lattice in the way that  preserves the star-triangle integrability, that is defining the cubic vertices of the theory corresponding to triangular faces in the lattice, and deriving all other interactions by star-triangle moves. Given the set of three lines that cross forming a triangle, the propagators in the corresponding cubic vertex transform under a representation $(\tfrac{a}{2},\tfrac{b}{2})$. Thus the propagator  is defined as follows:
\beq
\left[G_{\Delta}(x)\right]^{\boldsymbol{\alpha} \dot{\boldsymbol{\alpha}}}_{\boldsymbol{\dot \beta} \boldsymbol{\beta}} = \frac{(\boldsymbol{\sigma}^{\alpha_1 \dot{\alpha_1}}_{\mu_1}\otimes\cdots\otimes \boldsymbol{\sigma}^{\alpha_a \dot{\alpha_a}}_{\mu_a})  \hat x^{\mu_1}\cdots  \hat x^{\mu_a} \, (\bar{\boldsymbol{\sigma}}_{\dot{\beta_1}\beta }^{\nu_1}\otimes\cdots\otimes \bar{\boldsymbol{\sigma}}_{\dot{\beta_b}\beta_b}^{\nu_b})  \hat x_{\nu_1}\cdots  \hat x_{\nu_b} }{\left(x^2\right)^{\Delta}}\,,
 \label{spinpropagators}
\eeq
where $(\cdots)$ stands for symmetrysation over greek indices, which run over $1,2$ and $\hat x^{\mu} = x^{\mu}/|x|$. We shall use the compact notation 
\beq
\left[G_{\Delta}(x)\right]^{\boldsymbol{\alpha} \dot{\boldsymbol{\alpha}}}_{\boldsymbol{\dot \beta} \boldsymbol{\beta}} = \left(x^2\right)^{-\Delta} \left[ \boldsymbol{x} \right]^{\boldsymbol{\alpha} \dot{\boldsymbol{\alpha}}} \left[\bar{\boldsymbol{x}} \right]_{\dot{\boldsymbol{\beta}}\boldsymbol{\beta}} = \left(x^2\right)^{-\Delta} \left[ \boldsymbol{x} \right]^{a} \left[\bar{\boldsymbol{x}} \right]^{b}\,,
\eeq
in order to give the analytic expression of the star-triangle identity of figure \ref{STR_ker}:
\begin{multline}\label{STR_ker}
\int d^4 y\,\frac{[\mathbf{\overline{x-y}}]^{2s_3}
[\mathbf{x-y}]^{2s_1}}{(x-y)^{2 (u+2)}}
\frac{[\mathbf{ \overline{y}}]^{2s_1}\mathbf{R}_{2s_1 2s_2}(u+v)
[\mathbf{y}]^{\ell}}{y^{-2(u+v)}}
\frac{[\mathbf{\overline{y-z}}]^{2s_2}
[\mathbf{y-z}]^{2s_3}}{(y-z)^{2 (v+2) }} = \\ = c_{s_1,s_2,s_3}(u,v)\times
\frac{[\mathbf{ \overline{x}}]^{2s_3}\mathbf{R}_{2s_2,2s_3}(v)
[\mathbf{x}]^{2s_2}}{x^{-2 v}}
\frac{[\mathbf{\overline{x-z}}]^{2s_2}
[\mathbf{x-z}]^{2s_1}}{(x-z)^{2(u+v+2)}}
\frac{[\mathbf{ \overline{z}}]^{2s_1}\mathbf{R}_{2s_1,2s_3 }(u)
[\mathbf{z}]^{2s_3}}{z^{-2u}}\,,
\end{multline}
where $c_{s_1,s_2,s_3}(u,v)$ is a simple ratio of $\Gamma$-functions of $u$ and $v$ , and $s_1,s_2,s_3$ are the spin of each of the three intersecting slopes (for details see formula (2.31) in \cite{Derkachov:2021rrf}).

\begin{figure}[H]
\begin{center}
\includegraphics[scale=1.4]{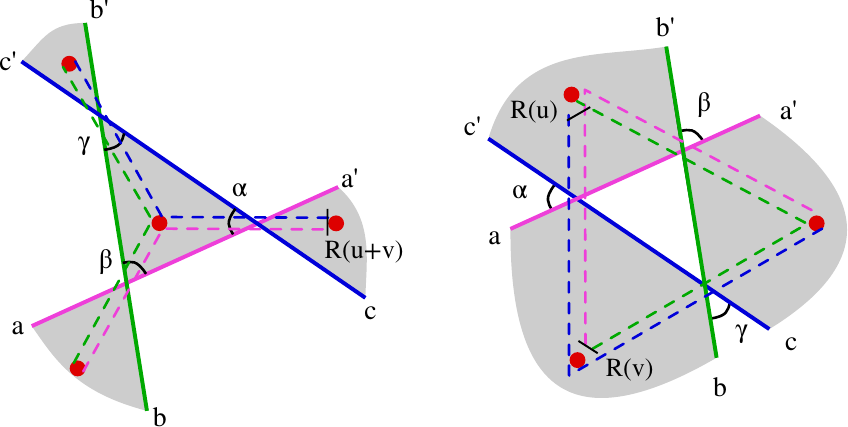}
\end{center}
\caption{\textbf{Left:} Diagrammatic form of the l.h.s. \eqref{STR_ker} of the star-triangle duality as a \emph{star} of propagators. \textbf{Right:} Diagrammatic form of the r.h.s. \eqref{STR_ker} of the star-triangle duality as a \emph{triangle} of propagators. The spinor indices are mixed by the $SU(2)$ fused $\mathbf{R}$-matrices that are contracted with the matrices $\boldsymbol{\sigma}, \boldsymbol{\overline{\sigma}}$ appearing in the definition \eqref{spinpropagators}, and their position is denoted by short segments. The angles are $\alpha= \pi (u+v+2)/2,\,\beta = - \pi v/2,\, \gamma = - \pi u/2 $. Magenta, blue and green lines are associated with spins $s_1,s_2$ and $s_3$ respectively.}
\label{braid_STR}
\end{figure}

\begin{figure}[h]
\begin{center}
\includegraphics[scale=0.9]{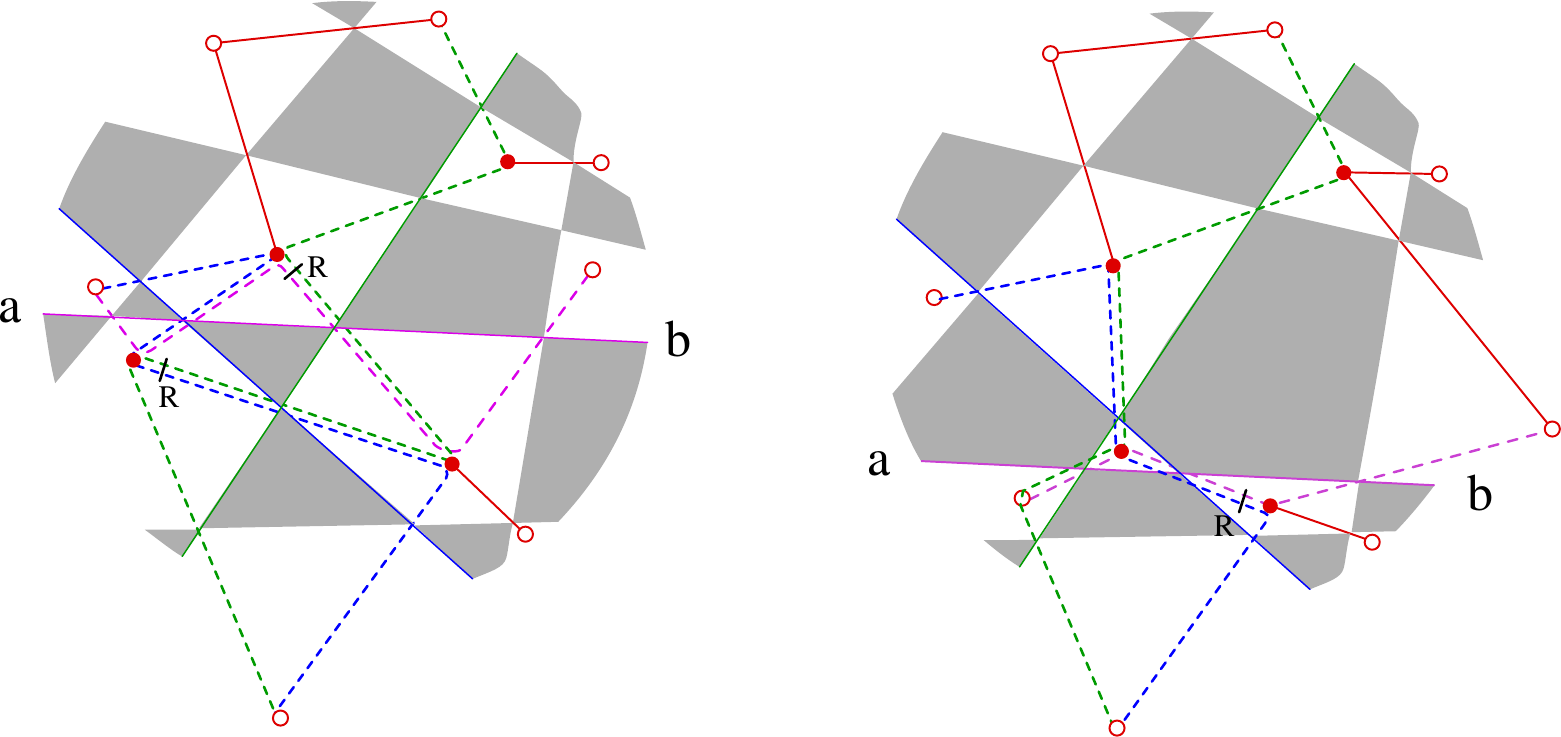}
\caption{The analogue of figure \ref{loom_circle} in presence of spinning Baxter lattice/spinning propagators. Coloured solid lines of the Baxter lattice carry a certain $SU(2)$ symmetric representation: different colors depict, in general, different spins. The propagators \eqref{spinpropagators}  that cross a lattice vertex featuring one (two) coloured lines are dressed with one (two) numerators in the corresponding $SU(2)$ symmetric representation(s). Solid lines are scalar propagators, while dashed lines indicate the Pauli matrix structure(s) in the numerators of \eqref{spinpropagators}.}
\label{loom_fermions}
\end{center}
\end{figure}
The matrices  $\mathbf{R}_{ab}(u)$ are solutions of the Yang-Baxter equation acting on the tensor product of $(a+1)$- and $(b+1)$-dimensional symmetric irrep of $SU(2)$. They can be obtained via fusion procedure \cite{Kulish:1981gi}. In contrast to  the scalar Loom FCFTs, in presence of fermions the loom moves change the configuration of such $\mathbf{R}$-matrices, besides producing a factor with a ratio of $\Gamma$-functions. Nevertheless, the integrability of Feynman integrals -- thus of the perturbative expansion of correlators -- is preserved thanks to Yang-Baxter property. The derivation of \eqref{STR_ker} is indeed a  straightforward generalisation of the scalar case. The details of it can be found  in \cite{Derkachov:2021rrf}.
An example of such a theory including the spinor fields is  the so-called $\mathcal{N}=2$ FCFT \cite{Pittelli:2019ceq}
\begin{equation}
\label{N2_fish}
\mathcal{L}_{\mathcal{N}=2} = N \,\text{Tr}\left[\bar \lambda_1 (\boldsymbol{\sigma} \cdot \partial) \lambda_1+\bar \lambda_2 (\boldsymbol{\sigma} \cdot \partial) \lambda_2 + 4\pi i \xi \left( \bar \phi \Delta \phi +   \lambda_1 \bar \phi \lambda_2 +\bar \lambda_1 \phi\bar  \lambda_2  \right) \right]\,.
\end{equation}
In  the framework of Loom FCFT, it corresponds to a special reduction of $M=3$ slopes in $d=4$ dimensions, with angles 
\begin{equation}
\theta_{12}= \theta_{23}=\theta_{31} = \frac{2\pi}{3}\,,
\end{equation}
and spins $s_1=s_2=0$ and $s_3=\tfrac{1}{2}$.
Triangular faces in the Baxter lattice correspond to the cubic vertices \eqref{cubic} where now the fields $v$ and $w$ have scaling dimension $\Delta'_{23}=\Delta'_{31}=\tfrac{3}{2}$ and transform in the representations $(\tfrac{1}{2},0)$ and $(0,\tfrac{1}{2})$ under spacetime rotations, namely  as  left and right Weyl fermions. The field $u$ is scalar, and it has the scaling dimension $\Delta_{12}'= 1$.
\begin{figure}[h]
\begin{center}
\includegraphics[scale=1]{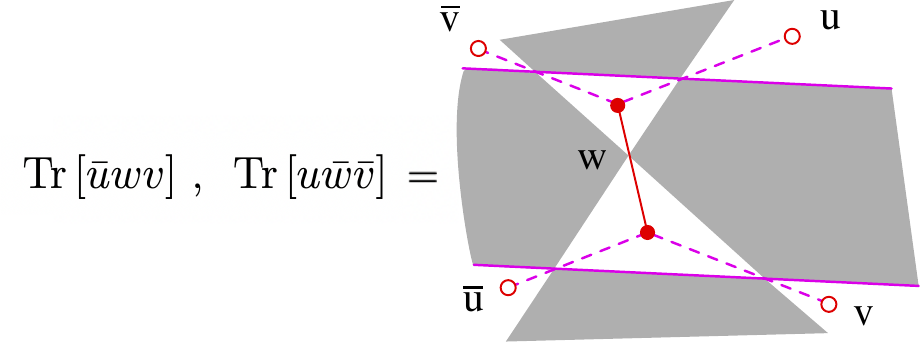}
\caption{Two couple of cubic vertices $\text{Tr}\left[\bar u v w\right]$ and $\text{Tr}\left[ u\bar  v\bar  w\right]$ in presence of a spinning slope (magenta) in the Baxter lattice.}
\label{fig:loom_f_3}
\end{center}
\end{figure}

\begin{figure}[h]
\begin{center}
\includegraphics[scale=1]{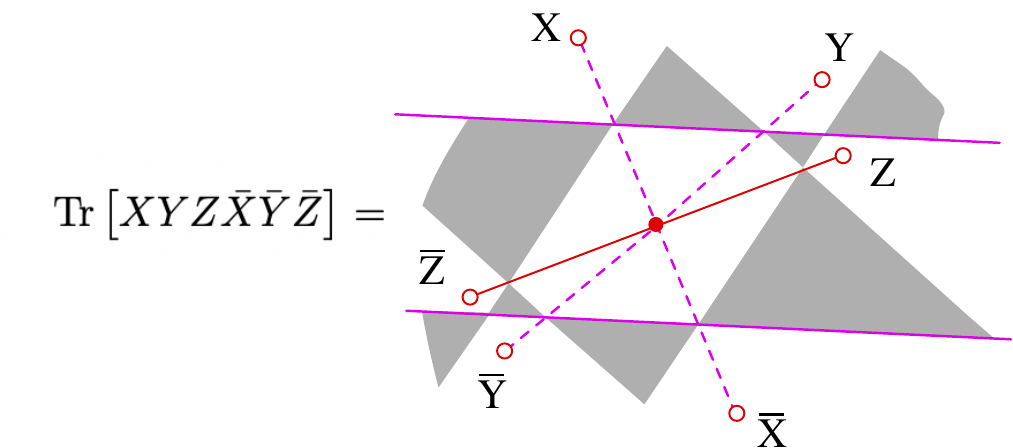}
\caption{The sextic vertex of $3$-Loom FCFT in presence of a spinning slope (magenta) in the Baxter lattice.}
\label{fig:loom_f_6}
\end{center}
\end{figure}

With the identification $\lambda_1=v$, $\lambda_2=w$ and $\phi=u$ we recover from \eqref{cubic} the original notation of \eqref{N2_fish}. We show this vertex in Fig.~\ref{fig:loom_f_3} using the notation of Fig.~\ref{loom_fermions}.
The interaction in \eqref{N2_fish} is only the cubic sector of $M=3$ Loom FCFT, obtained whenever all the other couplings are set to zero. The general $3$-Loom FCFT with spins $(0,0,\tfrac{1}{2})$ includes also the interactions \eqref{sextic}-\eqref{quartic}. For instance, the highest valency vertex is now a sextic interaction of two right fermions (duals to $u,\bar u$), two left fermions (duals to $v,\bar v$) and two scalars (duals to $w,\bar w$). In the notation of Fig.~\ref{loom_fermions}, we can draw the sextic interaction as presented on Fig.~\ref{fig:loom_f_6}.
More involved examples can feature more than one spin structure in the theory, with two or more spinning slopes in the Baxter lattice. In such a case we have  to deal with vertices having also an $SU(2)\otimes SU(2)$ tensor structure that mixes spin degrees of freedom via the  fused $\mathbf{R}$-matrix, constrained by integrability.

\section{Loom CFT amplitude as a sum over Loom graphs} 

The fishnet amplitudes with disc topology at \(N=\infty\) have been introduced and studied for the bi-scalar FCFT~\cite{Chicherin:2017cns,Chicherin:2017frs,Korchemsky:2018hnb, Chicherin:2022nqq}. It was observed that such an amplitude is always given by a single graph cut out of the regular square lattice. We will generalise here these amplitudes to any $M$-Loom FCFT and show that only a very limited number of diagrams contributes to such amplitude, all related by star-triangle moves.  

Let us argue that in the large-$N$ limit of any Loom FCFT the scattering amplitudes with disk topology are described by a finite set of processes, i.e. of Feynman integrals, at any couplings. 
Following the LSZ prescription, a scattering amplitude with $n$ particles and disk topology is the residue on-shell $p_k^2=0\,,\,k=1,\dots, n$, of the Fourier transform of an $n$-point single-trace correlator, for instance
\begin{equation}
\label{traceamp}
\text{Tr}\left[\Phi_{12}(x_1) \phi_{23}(x_2) \bar{\Phi}_{12}(x_3) \cdots \bar{\phi}_{46}(x_n) \right]\,.
\end{equation}
The weak-coupling expansion of such correlator in the planar limit, with the disc topology, has a finite number of contributions that can be readily read  out of the Baxter lattice. In fact, any such Feynman diagram can be obtained starting from a very large and general Baxter lattice by drawing a closed path that cuts the propagators of the fields $\Phi_{12}\,,\, \phi_{23}\,,\, \bar{\Phi}_{12}\,, \dots \bar{\phi}_{46}$ in a given order. Once such diagram is computed, all other contributions to the same correlator are obtained by a number of star-triangle transformations, that is of moves of lines on the Baxter lattice, with fixed order of external legs on the boundary. There exists clearly only a finite set of such moves: the related diagrams, say $B$ and $B'$ are related to each other by a chain  of the star-triangle moves. Then according to the formula \eqref{STR} dey differ by a finite ratio $\mathcal{V}(B\to B')$ of $\Gamma$-functions, i.e.
\begin{equation}
B(x_1,\dots, x_n) = \mathcal{V}(B\to B')B'(x_1,\dots, x_n)\,,
\end{equation}
and they carry in general a different set of couplings.
\begin{figure}
\begin{center}
\includegraphics[scale=0.6]{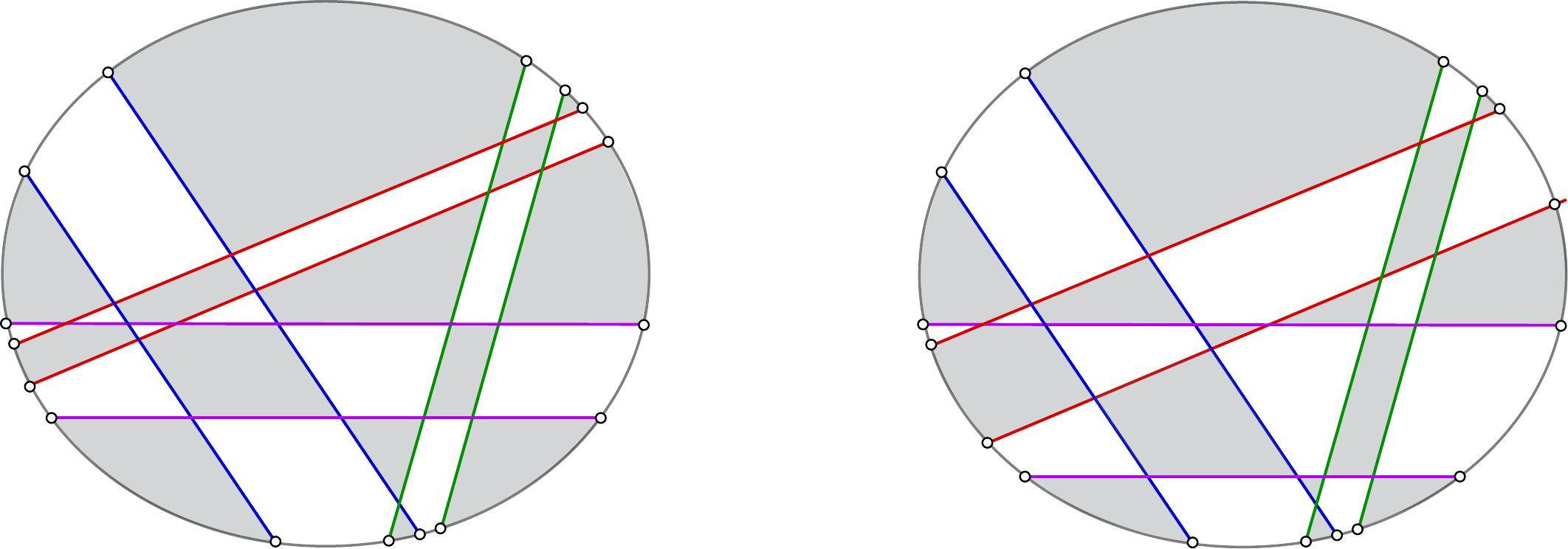}
\caption{Two possible  Baxter lattices dual to the Feynman diagrams contributing to a certain $8$-particles amplitude involving four different slopes. They are related via star-triangle moves.
}
\label{ampli2}
\end{center}
\end{figure}
We can define systematically a standard ``base" diagram $B_0$ that defines the equivalence class $[B_0]$ of all the other diagrams that are related to it by star-triangle moves. For a given $n$-point diagram its base diagram is defined as follows: let us consider the dual portion of Baxter lattice and let the number $m(h)$ be the number of times the slope $h$ occurs in it. Let us draw a bundle of $m(h)$ lines with slopes $h$ for each of the slope that appears in the given diagram. The bundles can be ordered so that they form a polygon with at most $M$ edges, since $1\leq h \leq M$. Finally, the base diagram is obtained by cutting a close contour out of these bundles, such that to reproduce the boundary conditions, i.e. the order of fields inside the trace \eqref{traceamp}. An example is given already at the level of Baxter lattice by figure~\ref{ampli2}: the portion of lattice on the left is associated to its base representative $B_0$ on the right. Finally, the full answer for a given planar correlator \eqref{traceamp} can be arranged as the following sum
\begin{equation}
\sum_{B \in [B_0]} (\text{coupling of }B ) \times B(x_1,\dots, x_n) = B_0(x_1,\dots, x_n)  \sum_{B \in [B_0]} \mathcal{V}(B_0 \to B) (\text{coupling of }B )\,  
\end{equation}
that is a homogeneous polynomial in the couplings that multiplies a single function of the coordinates. Here the (coupling~of~\(B\)) means the product of all single trace couplings of the graph $B$.
It was shown in~\cite{Chicherin:2017cns,Chicherin:2017frs}  that in the bi-scalar theory~\eqref{FCFTbiscalar} the correlators of the type~\eqref{traceamp}, defining the off-shell bi-scalar amplitudes, obey the Yangian symmetry: a certain ``lasso" operator can be applied at the external coordinates of such  correlator with disc topology, for which this correlator appears to be an eigenfunction.\footnote{ We recall that this symmetry develops an anomaly on-shell, due to the region where the loop momenta are collinear with external ones \cite{Chicherin:2022nqq}, so the Yangian symmetry applies only to the off-shell amplitudes.} It is highly probable that the same Yangian symmetry is proper to arbitrary correlators of the type~\eqref{traceamp} in any Loom FCFT.

\section{``tT" invariance and reality properties  of the spectrum}
\label{sec:tTsym}
As it was noticed already in the early papers on Fishnet CFTs~\cite{Gurdogan:2015csr,Caetano:2016ydc,Gromov:2017cja,Grabner:2017pgm,Kazakov:2018qbr,Gromov:2018hut}, the spectrum of their anomalous dimensions of consists of either positive or complex conjugate dimensions.  Such is the case  of $d$-dimensional bi-scalar Fishnet CFT~\eqref{FCFTbiscalar}.

The appearance of complex dimensions is a ubiquitous feature in this  non-unitary CFT, but the fact that they enter in complex-conjugate pair needs some explanation. Let us discuss the symmetry behind this phenomenon. 

Consider first the case of FCFT~\eqref{FCFTbiscalar}. If we apply the time reversal,  $T$ transformation. It corresponds to the hermitian conjugation to this Lagrangian: $
{\cal L}(X,Y)\,\to\, \overline{{\cal L}(X,Y)}
$.  Under this transformation, both kinetic terms do not change, whether as the last, interaction term transforms into its ``chiral" dual $(4 \pi)^{d/2} \xi^2\,\, \text{Tr} \left[XY  \bar X\bar Y\right]$. 
  If we than apply the transposition: $X,Y\to(X)^t,(X)^t$, called here as~$t$ transformation, the matrix fields inside the trace will be reordered in the opposite way, thus rendering the original, chiral interaction terms:
\begin{align*}
\text{Tr}\left[XY\bar X\bar Y\right]\quad \xrightarrow{{T}} \quad \overline{\text{Tr}\left[XY\bar X\bar Y\right]}\quad=\quad \text{Tr}\left[ \bar X^t  \bar Y^t  X^t  X^t \right] \quad\xrightarrow{{t}} \quad \text{Tr}\left[\bar X\bar Y X Y\right]
\end{align*}
 That means that the composition of two transformations, which we call here $tT$ transformation, is a symmetry of this Lagrangian. Since the functional measure is also obviously $tT$ invariant,  this is the symmetry of the whole theory. 

What are the consequences of this symmetry on the spectrum of anomalous dimensions? It is clear that the operators are not necessarily $tT$-symmetric, e.g. 
\begin{align*}
\text{Tr}[XXY\bar X]\quad\overset{tT}\to\quad \text{Tr}[\bar X\bar X\bar YX]\,.
\end{align*}
Then the two point function of such an operator, defining its conformal dimension, $tT$  transforms in the following way:
\begin{align*}
\left[\langle \bar{\cal O}(x) {\cal O}(0)\rangle\right]^{{\rm tT}}\,=\,\langle \bar{\cal O}^{{\rm tT}}(x) {\cal O}^{{\rm tT}}(0)\rangle\,=\,  |x|^{-2\Delta^*},
\end{align*}
which means that the conformal dimensions are either equal and real for both,  ${\cal O}$ and ${\cal O}^{tT}$ operators, or they are complex conjugate.   

However, there are a few subtleties in the interpretation of such behavior of anomalous dimensions. Take as an example the spectrum of the length-2 exchange operators in the 4-point function 
\begin{align}
\label{K_X}
& K_X(x_1,x_2;x_1',x_2')=\frac{1}{N}\langle \Tr[X(x_1)X(x_2)] \Tr[\bar X(x_1')\bar X(x_2')]\rangle
\,,
\end{align}
in the FCFT~\eqref{FCFTbiscalar}. The direct calculation generalizing the eq.(6) of~\cite{Kazakov:2018qbr}  to any $\omega$ gives the following equation for the spectrum of such operators:
\begin{align}
\label{exchangeDelta}
    \frac{\Gamma \left(\frac{d}{2}-\omega \right)^2 \Gamma \left(\frac{\Delta +S}{2}+\omega \right) \Gamma \left(\frac{d}{2}+\omega -\frac{\Delta-S }{2}\right)}{\Gamma (\omega )^2 \Gamma \left(\frac{\Delta +S}{2}-\omega \right) \Gamma \left(\frac{d}{2}-\omega -\frac{\Delta -S}{2}\right)}=\xi^{4}
\end{align}
In order to demonstrate here the issues of reality/complexity of the spectrum at various values of $\omega$ (i.e. for various angles $\alpha=\frac{2\pi}{d}\omega$ of the 2-loom), we will consider, for the sake of simplicity of solutions for $\Delta(\xi)$, the case of $d=4$ and of the spin $S=0$.

At $\omega=1/2$ \eqref{exchangeDelta} reduces to $\Delta ^2+16 \xi ^4+3=4 \Delta$, so that the two real solutions have the following perturbative expansions 
\begin{align}
 \Delta_1&=1+8 \xi ^2+32 \xi ^4+256 \xi ^6+2560 \xi ^8+O\left(\xi ^9\right) \notag\\
  \Delta_2&=3-8 \xi ^2-32 \xi ^4-256 \xi ^6-2560 \xi ^8+O\left(\xi ^9\right)
\end{align}
 Notice that the second one is the ``shadow" solution, so we are left only with the first one in the spectrum, corresponding to the exchange operator $\Tr[X^2(x)]$. 
 
The case  $\omega=3/2$, which corresponds to the exchange operator $\Tr[Y^2(x)]$, has quite similar reality properties. The spectral equation \eqref{exchangeDelta} reduces now to $- (\Delta -5) (\Delta -3)^2 (\Delta -1)^2 (\Delta +1)=256 \xi ^4$, so that we have 6 real solutions with the following perturbative expansions 
\begin{align}
 \Delta_1&=-1+\frac{2 \xi ^4}{3}+\frac{20 \xi ^8}{27}+\frac{647 \xi ^{12}}{486}+O\left(\xi ^{15}\right) 
 \notag\\
  \Delta_2&=5-\frac{2 \xi ^4}{3}-\frac{20 \xi ^8}{27}-\frac{647 \xi ^{12}}{486}+O\left(\xi ^{15}\right)
 \notag\\
\Delta_3&=1+2 \sqrt{2} \xi ^2+3 \xi ^4+\frac{53 \xi ^6}{4 \sqrt{2}}+30 \xi ^8+\frac{20679 \xi ^{10}}{128 \sqrt{2}}+\frac{1821 \xi ^{12}}{4}+\frac{5543373 \xi ^{14}}{2048 \sqrt{2}}+O\left(\xi ^{15}\right)
 \notag\\
\Delta_4&=3-2 \sqrt{2} \xi ^2-3 \xi ^4-\frac{53 \xi ^6}{4 \sqrt{2}}-30 \xi ^8-\frac{20679 \xi ^{10}}{128 \sqrt{2}}-\frac{1821 \xi ^{12}}{4}-\frac{5543373 \xi ^{14}}{2048 \sqrt{2}}+O\left(\xi ^{15}\right)
  \notag\\
\Delta_5&=1-2 \sqrt{2} \xi ^2+3 \xi ^4-\frac{53 \xi ^6}{4 \sqrt{2}}+30 \xi ^8-\frac{20679 \xi ^{10}}{128 \sqrt{2}}+\frac{1821 \xi ^{12}}{4}-\frac{5543373 \xi ^{14}}{2048 \sqrt{2}}+O\left(\xi ^{15}\right)
  \notag\\
\Delta_6&=3+2 \sqrt{2} \xi ^2-3 \xi ^4+\frac{53 \xi ^6}{4 \sqrt{2}}-30 \xi ^8+\frac{20679 \xi ^{10}}{128 \sqrt{2}}-\frac{1821 \xi ^{12}}{4}+\frac{5543373 \xi ^{14}}{2048 \sqrt{2}}+O\left(\xi ^{15}\right)
\end{align}
we see that, again, all solutions are real at real $\xi^2$. The first solution corresponds to ``shadow" operator and should be dropped from the spectrum.  

If now we consider $\omega=1$ -- the ``isotropic"  biscalar FCFT coming from ${\cal N}=4$  SYM~\cite{Gurdogan:2015csr} -- we obtain the 4th order equation $(\Delta -4) (\Delta -2)^2 \Delta =16 \xi ^4$ with the following 4 perturbative solutions
\begin{align}
 \Delta_1&=2-2 i \xi ^2+i \xi ^6-\frac{7 i \xi ^{10}}{4}+\frac{33 i \xi ^{14}}{8}+O\left(\xi ^{18}\right)
 \notag\\
  \Delta_2&=2+2 i \xi ^2-i \xi ^6+\frac{7 i \xi ^{10}}{4}-\frac{33 i \xi ^{14}}{8}+O\left(\xi ^{18}\right)
 \notag\\
\Delta_3&=-\xi ^4+\frac{5 \xi ^8}{4}-\frac{21 \xi ^{12}}{8}+\frac{429 \xi ^{16}}{64}+O\left(\xi ^{18}\right),
 \notag\\
\Delta_4&=4+\xi ^4-\frac{5 \xi ^8}{4}+\frac{21 \xi ^{12}}{8}-\frac{429 \xi ^{16}}{64}+O\left(\xi ^{18}\right)
\end{align}
Here we have two real solutions, out of which the $\Delta_3$ corresponds to the ``shadow" operator of the last one. But we also get 2 complex conjugate solutions. Their appearance is due to the quadratic cut w.r.t. $\xi^4$ in the spectral equation~\eqref{exchangeDelta} at $S=0,\,\, \omega=d/4$, at any $d$.   

Actually, one of these two complex dimensions should be also considered as of the ``shadow''  operator. Indeed, it is know~\cite{Grabner:2017pgm,Gromov:2018hut} that  the correlators of bi-scalar theory at $\omega=d/4$ develop a cut at  $\xi^2\in(0,\infty)$, due to the presence of two complex conjugate double-trace terms~\eqref{XXbXbX}. The coupling for these terms has two complex conjugate critical  values (functions of $\xi$ given by zeros of the $\beta$-function of the type~\eqref{alphaXX}). To regularize the theory we have to add a little complex part to $\xi^2\to \xi^2\pm i\epsilon$. Then the theory, depending on the sign will run in the IR to one of these critical points. For plus sign, the eigenvalue $\Delta_1$ corresponds to the ``shadow" operator and $\Delta_2$ - to the physical one, and vice verse for the minus sign. In conclusion, our theorem of positivity of spectrum is valid up to the fact that some complex dimensions can correspond to shadow operators and thus they are not in the spectrum.  The last phenomenon happens for particular angles of the loom, such as $\alpha=\frac{2\pi}{d}\omega=\pi/2$ in the example above.

We see similar patterns of reality in   the behavior of two dimensions on Fig.2 of~\cite{Gromov:2017cja}: for small enough coupling  $\xi$, the operator $\tr X_1^3$ has a real dimension. However, at the value $\xi^3\simeq0.21$  it collides with its ``shadow"
 and they form a complex conjugate pair.
A similar picture is seen in analytic perturbative calculations ~\cite{Gromov:2017cja}, (e.g. the eqs.(7.11-13) therein), but both complex conjugated pairs of dimensions can correspond to the physical operators.   
 \footnote{We are grateful to Gregory Korchemsky for very pertinent comments on the violation of $tT$-symmetry.}

The $tT$  symmetry is a common feature  not only of this bi-scalar fishnet CFT but it is proper to any generalised, Loom FCFT at generic values of the angles in the Baxter lattice. Indeed, all  kinetic terms of any Loom FCFT are obviously $tT$~symmetric, which is also true for the single-trace  interaction terms (see e.g.\eqref{sextic}-\eqref{cubic} for the case of $M=3$ slopes). Due to the chirality of the Lagrangian, the $T$ transformation changes all single-trace  chiral terms to anti-chiral and replaces their couplings with the complex conjugates, whether as the $t$~transformation sets them back to the original ordering, but also replaces each field with its hermitian conjugate. Due to the structure of the Baxter lattice, the vertices in the Loom FCFT always appear in pairs \eqref{conjugate_Vertices}, so that the single-trace interactions of the FCFT Lagrangian are $tT$-invariant.\footnote{We can imagine a  chiral FCFT in terms of planar Feynman graphs drawn on a transparent glass: looking at them from behind the glass we see the planar graphs of anti-chiral FCFT.}  The double-trace terms described in the section~\ref{sec:LoomFCFT} have the same \(tT\)-symmetry since they have the same sequence of fields as in single-trace vertices but split into two traces, and their coupling are real, as demonstrated in section~\ref{sec:renorm} and discussed in the last paragraph of section \ref{renom_3} regarding the $3$-Loom FCFT.   Therefore, the spectrum of  Loom FCFT at any \(M\) and generic angles obeys  the same reality property: it contains either real conformal dimensions or complex conjugate pairs. However, as it was demonstrated above for the bi-scalar model at $\omega=d/4$, for particular angles of the Baxter lattice, new double-trace terms with complex couplings can appear which will violate the $tT$-symmetry, and hence the reality property of a part of the spectrum.

This $tT$ symmetry reminds the $PT$~symmetry  of non-unitary quantum mechanical systems and QFTs  proposed in~\cite{Bender:2007nj,Bender:1998ke,Ai:2022csx}, where the spectrum of energies obeys the same properties: it contains either real or complex conjugate energy levels. The latter usually appear for certain critical finite values of couplings and their appearance is interpreted as the breakdown of $PT$~symmetry.

\section{Conclusions}
\label{sec:Conclusions}

In this work, we constructed the most general family of Fishnet CFTs --  the conformal theories of matrix fields with ``chiral" planar interactions in various dimensions. The distinguished feature of these Loom FCFTs is that their perturbation theory contains only a very limited set of Feynman diagrams, dual to the so called Baxter lattices -- a set of intersecting straight lines on the plane. These diagrams appear to be integrable, as was noticed long ago by A.~Zamolodchikov~\cite{Zamolodchikov:1980mb}: all such Feynman graphs at a given order of perturbation theory are related by star-triangle identities, so that effectively only one or a handful of graphs are subject to non-trivial computations.
Particular cases of such FCFTs have been proposed in the past, starting from the bi-scalar FCFT~\cite{Gurdogan:2015csr,Kazakov:2018qbr} appearing in \(\gamma\)-twisted \(\mathcal{N}=4\) SYM reduction,  or \(3d\)  FCFT~\cite{Caetano:2016ydc} --  a reduction of \(\gamma\)-twisted ABJM model. 

Our present work gives the general scheme of construction of such theories, integrable in 't~Hooft limit, in any dimension and for most general dimensions of matrix quantum fields. Each such  FCFT is characterised by  Baxter lattices consisting of a system of \(M\) crossing straight lines with different slopes, and their parallels. This lattice serves as a the ``loom" for ``weaving" the dual integrable planar Feynman graphs. That's why we call such a theory the $M$-Loom FCFT. Its action has \(M(M-1)\) matrix scalar fields,  and a number of scalar interactions, quickly (exponentially) increasing with \(M\). On the other hand, one can choose for particular physical applications a particular subset of non-zero couplings, thus reducing the number of  interactions and decoupling certain fields.
We also outlined the generalisation of these \(M\)-Loom FCFTs to spinning matrix fields in $d=4$ dimensions. 

Let us list  some of the prospects and applications concerning our construction of \(M\)-Loom FCFTs:

\begin{itemize}
\item  \(M\)-loom FCFTs
 are non-unitary, logarithmic CFTs, with the non-diagonalisable mixing matrix of conformal operators containing Jordan cells.  It would be interesting to understand the general structure of the spectrum of anomalous dimensions and of the OPE in these theories, to compute, or may be even to bootstrap the structure functions, along the lines of~\cite{Cavaglia:2021bnz}. 
\item  \(M\)-loom FCFTs obey a certain discrete symmetry which we called \(tT\) symmetry: it combines the \(T\)-transformation (time reversal) and the \(t\)-transformation (transpose of fields). Due to that, the conformal dimensions in the theory are real or enter into the spectrum in complex conjugate pairs. The property is similar to that of the spectrum of \(PT\)-symmetric non-unitary QM systems considered in~\cite{Bender:1998ke,Bender:2007nj}.  It would be interesting to push further this analogy and may be also to search for applications to condensed matter physics.

\item\ There is still no understanding how the integrability works for the so-called dynamical fishnet CFT -- the three-coupling reduction of \(\gamma\)-twisted \(\mathcal{N}=4\) SYM~\cite{Gurdogan:2015csr,Kazakov:2018gcy}. It seems that we cannot include it into our general \(M\)-loom FCFT scheme based on the star-triangle identity. Does it exist a generalisation of our construction, that likely relies on integrability with long range interactions, which is suitable for the dynamical fishnet?

\item\ The Fishnet CFT is a step in understanding the integrability of the full \(\mathcal{N}=4\) SYM. However, we still don't know how to construct the corresponding 1+1-dimensional \(PSU(2,2|4)\) spin chain (certainly with non-local interactions!) whose solution would be equivalent to the summation of all planar Feynman graphs of the theory. It may be that such graphs have a (well) hidden dynamical lattice structure, just a bit more complicated than the dynamical graphs of FCFT~\cite{Gurdogan:2015csr,Kazakov:2018gcy}.

\item\ The discovery of FCFTs triggered the study of the spectrum of logarithmic CFTs, in particular of its structure in terms of Jordan cells, starting from some toy-models called Hyper-Eclectic and Eclectic spin chains~\cite{Ahn:2020zly, Ahn:2021emp, Ahn:2022snr}. These have similar properties of the spectrum: the Hamiltonian is not diagonalisable, and its Jordan form contains Jordan cells. Their classification  is a problem of combinatorics of random walks (also known as ``stampedes" \cite{Olivucci:2021pss}), expressible in the language of Young tableaux. What is the general method for counting stampedes on a $M$-Loom diagram rather than on a bi-scalar one?

\item\ The study of the four-point Basso-Dixon Feynman integrals in the bi-scalar FCFT in four and two dimensions \cite{Basso:2017jwq, Derkachov:2018rot,Basso:2021omx, Kostov:2022vup} and later at any~$d$ \cite{Derkachov:2021ufp} and the use of Yangian symmetry to bootstrap other higher-point fishnet integrals for the planar amplitudes and correlators \cite{Chicherin:2017cns,Chicherin:2017frs,Loebbert:2019vcj, Corcoran:2021gda} has uncovered an amazingly rich structure. Indeed, it  allowed to make contact with the hexagonalisation techniques inspired by $AdS/CFT$ holography. It also shed some light on the problem of understanding the space of functions of higher-point higher-loop Feynman integrals \cite{Duhr:2022pch,Gurdogan:2020ppd}. The computation of similar quantities in other FCFTs from the Loom could provide  great results in this direction.

\item\ The study of the spectrum of FCFTs is a very tricky problem, in spite of the integrability. It involves the construction of a framework for the underlying non-compact spin chains, with the spins in principal series representations of conformal groups, analogous to the Algebraic Bethe Ansatz and inspired by the $SL(2,\mathbb{C})$ techniques of~\cite{Derkachov:2001yn,DeVega:2001pu}. The solution might have the ultimate formulation in terms of the Quantum Spectral Curve formalism, similar to that of the   \(\mathcal{N}=4\) SYM~\cite{Gromov:2013pga,Gromov:2014caa,Kazakov:2015efa}, based on the the Hasse diagram, \(QQ\)-relations for Baxter's \(Q\)~functions and on their analyticity properties. This construction was achieved so far only partially~\cite{Gromov:2017cja,Gromov:2019jfh}. 
\end{itemize}

\section*{Acknowledgements}
We would like to thank M.~Alfimov, B.~Basso, G.~Ferrando, E.~Trevisani and especially G.~Korchemsky for discussions and comments. We appreciate the kind hospitality  of the Kavli Institute for Theoretical Physics of the University of California Santa Barbara, where a part of this work has been done during the ``Integrability in String, Field, and Condensed Matter Theory" 2022 program. This research was supported in part by the National Science Foundation under Grant No.~NSF~PHY-1748958. Research at the Perimeter Institute is supported in part by the Government of Canada through NSERC and by the Province of Ontario through MRI. 

\appendix

\section{Vertices of $M=4$ Loom FCTF}
\label{app:vertices}
We list in the following all the vertices generated by the Baxter lattice with $M=4$ slopes, according to the notations for fields of section \ref{sec:4loom}.
\begin{itemize}
\item The vertex of valence $8$:
\begin{center}
\includegraphics[scale=0.9]{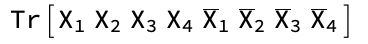}
\end{center}
\item The $8$ vertices of valence $7$:
\begin{center}
\includegraphics[scale=0.65]{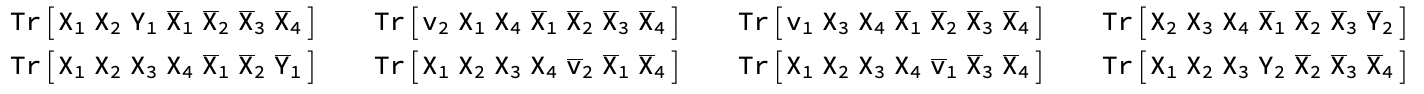}
\label{7-vertices}
\end{center}
\item The $28$ vertices of valence $6$:
\begin{center}
\includegraphics[scale=0.75]{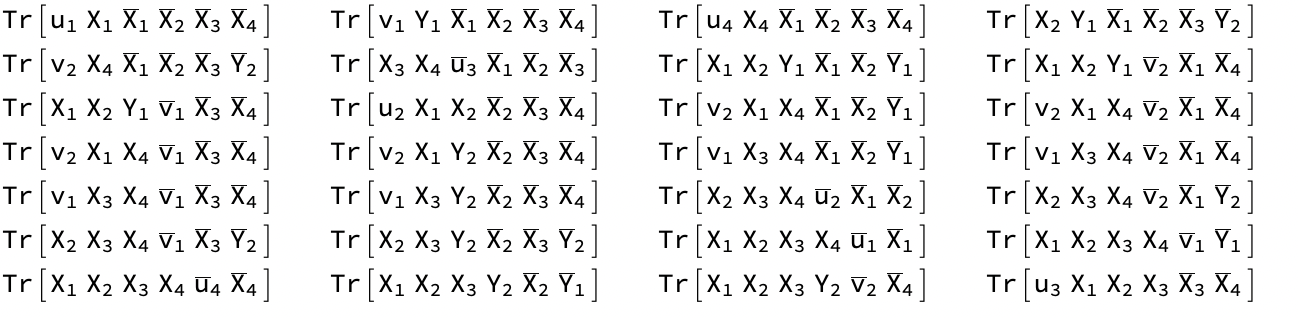}
\end{center}
\item The $48$ vertices of valence $5$:
\begin{center}
\includegraphics[scale=0.84]{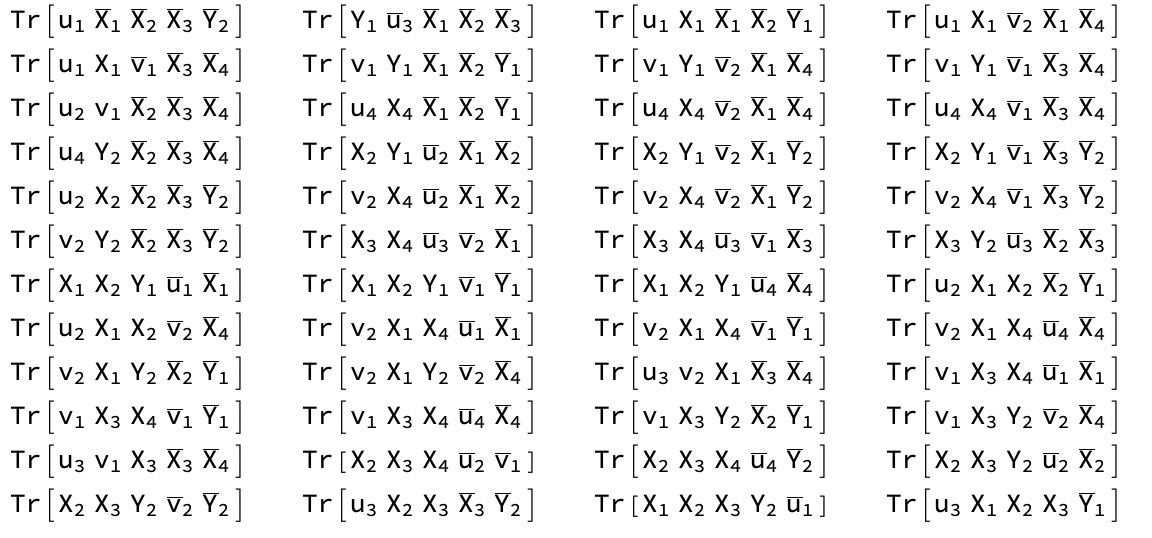}
\end{center}
\item The $38$ vertices of valence $4$:
\begin{center}
\includegraphics[scale=0.88]{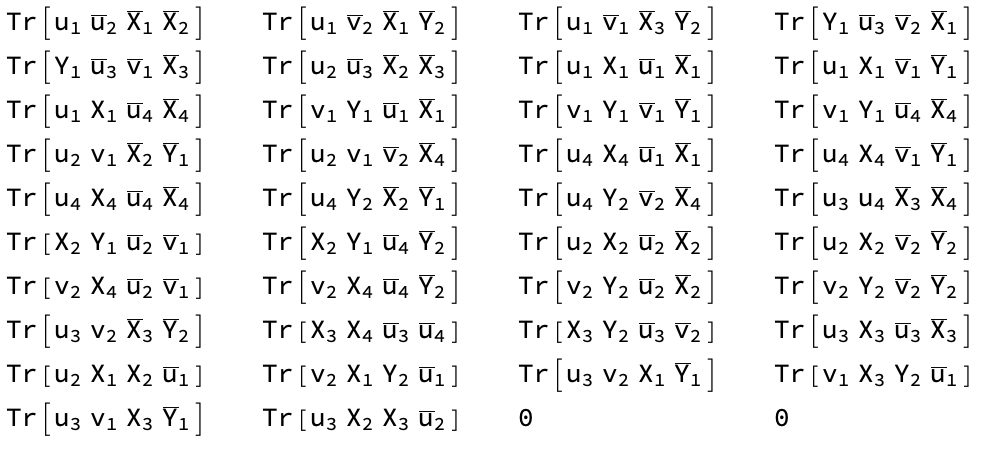}
\end{center}
\item The $8$ vertices of valence $3$:
\begin{center}
\includegraphics[scale=0.9]{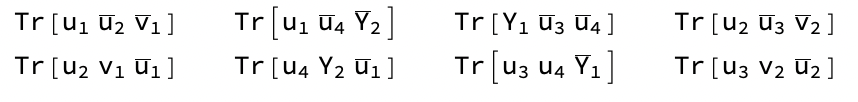}
\end{center}
\end{itemize}

\bibliographystyle{MyStyle}
\bibliography{Loom_FCFT}

\end{document}